\documentclass[nonblindrev]{informs3noheader}

\OneAndAHalfSpacedXI

\usepackage{natbib}
\bibpunct[, ]{(}{)}{,}{a}{}{,}%
\def\bibfont{\small}%

\usepackage[utf8]{inputenc}
\usepackage[english]{babel}
\usepackage{graphicx}
\usepackage{subcaption}
\newlength{\halfwidth}
\usepackage{color}
\usepackage{hyperref}
\usepackage{booktabs}
\usepackage{multicol}
\usepackage{xcolor}
\usepackage{graphicx, amsmath, amssymb, amssymb, amsfonts, enumerate, booktabs, float, adjustbox, hyperref, tikz, multirow,pgfplots,wrapfig,bm,bbm,graphicx}
\usepackage[titlenumbered,ruled]{algorithm2e}
\usepackage[flushleft]{threeparttable}

\allowdisplaybreaks

\definecolor{myblue}{RGB}{0,51,140}
\definecolor{myred}{RGB}{126,0,0}
\definecolor{mygreen}{RGB}{0,90,0}
\definecolor{mypurple}{RGB}{142, 36, 170}
\definecolor{mygray}{RGB}{200,200,200}

\usetikzlibrary{decorations.markings}
\usetikzlibrary{decorations.pathmorphing}
\usetikzlibrary{shapes.geometric, arrows,angles,quotes}
\usetikzlibrary{arrows,decorations.pathmorphing,backgrounds,positioning,fit,matrix}
\usetikzlibrary{shapes,decorations,arrows,calc,arrows.meta,fit,positioning}
\tikzset{
	circ/.style = {draw,circle,minimum size=2.5em,inner sep=1},
	ang/.style={draw, angle eccentricity=1.5,angle radius=0.4cm}
}

\newenvironment{edit}{\color{black}}
\MANUSCRIPTNO{}

\begin{document}

\ARTICLEAUTHORS{
	\AUTHOR{D. Bertsimas,
    V. Digalakis Jr.,
    A. Jacquillat,
    M. L. Li,
    A. Previero}
	\AFF{Sloan School of Management and Operations Research Center, Massachusetts Institute of Technology}
}
\RUNAUTHOR{Bertsimas, Digalakis Jr., Jacquillat, Li, Previero} 

\RUNTITLE{Where to locate mass vaccination sites?}

\TITLE{\Large Where to locate COVID-19 mass vaccination facilities?}

\vspace{-20pt}

\ABSTRACT{
	The outbreak of COVID-19 led to a record-breaking race to develop a vaccine. However, the limited vaccine capacity creates another massive challenge: how to distribute vaccines to mitigate the near-end impact of the pandemic? In the United States in particular, the new Biden administration is launching mass vaccination sites across the country, raising the obvious question of where to locate these clinics to maximize the effectiveness of the vaccination campaign. This paper tackles this question with a novel data-driven approach to optimize COVID-19 vaccine distribution. We first augment a state-of-the-art epidemiological model, called DELPHI, to capture the effects of vaccinations and the variability in mortality rates across age groups. We then integrate this predictive model into a prescriptive model to optimize the location of vaccination sites and subsequent vaccine allocation. The model is formulated as a bilinear, non-convex optimization model. To solve it, we propose a coordinate descent algorithm that iterates between optimizing vaccine distribution and simulating the dynamics of the pandemic. As compared to benchmarks based on demographic and epidemiological information, the proposed optimization approach increases the effectiveness of the vaccination campaign by an estimated 20\%, saving an extra 4,000 extra lives in the United States over a three-month period. The proposed solution achieves critical fairness objectives---by reducing the death toll of the pandemic in several states without hurting others---and is highly robust to uncertainties and forecast errors---by achieving similar benefits under a vast range of perturbations.

	}

	
\maketitle
	
\section{Introduction}

The outbreak of the COVID-19 pandemic has started a global race to develop vaccines, fueled by extensive investments, governmental support, and scientific breakthroughs. Thanks to these unprecedented efforts, the scientific community delivered the good news that the whole world was eagerly awaiting. By Summer 2020, several vaccines had been developed. By the end of 2020, several vaccines got approved for emergency use and hundreds more were going under development and testing. Whereas vaccine development used to take years and even decades, these results rank, with no doubt, among the greatest scientific achievements \citep{lurie2020developing,graham2020rapid}.

Unfortunately, discovering and developing a vaccine for COVID-19 was just the beginning---it will now take months to produce, distribute, and deliver vaccines at scale. The world has quickly come to the realization that vaccines cannot be made available immediately to everyone, and policy makers need to make tough decisions to pilot vaccine distribution. A global consensus has naturally emerged to prioritize to healthcare workers, other front line workers, and vulnerable populations such as older people and people with comorbidities \citep[see, e.g.,][]{national2020framework}. Within these general principles, each jurisdiction is designing more detailed eligibility guidelines to distribute vaccines effectively and equitably \emph{within a population}, based on demographic, clinical and geographic factors. However, a question remains open: how to plan vaccine distribution \emph{across populations}, that is, how to allocate a limited vaccine supply across communities, across provinces, and even across countries?

In the United States, this question gained prominence in the midst of a presidential transition. In particular, the new Biden administration relies on higher extents of federal coordination in vaccine distribution, as opposed to a more decentralized approach at the state level. In one of its first major decisions, the administration started opening mass vaccination sites, with many more planned over the next few weeks.\footnote{www.nbcnews.com/politics/white-house/federal-government-opening-first-mass-covid-19-vaccination-sites-california-n1256611} This environment raises the critical question of where to locate these vaccination sites. Obviously, these decisions need to adhere to a number of political and fairness considerations---most notably, there must be at least one site per state. Yet, there remains flexibility to use mass vaccination sites as a strategic lever to effectively combat the pandemic. 

This paper addresses this question with a novel data-driven approach, combining epidemiological modeling and prescriptive analytics, to optimize the location of vaccination sites and the subsequent allocation of vaccines. To this end, we leverage a recent compartmental epidemiological model called DELPHI (Differential Equations Lead to Predictions of Hospitalizations and Infections), which extends Susceptible-Exposed-Infected-Recovered (SEIR) models to capture critical drivers of the COVID-19 pandemic: (i) under-detection due to limited testing, (ii) governmental and societal response, and (iii) declining mortality rates \citep{li:20}. The DELPHI model has been fitted from historical data at the country level, at the state level in the United States, and at the province level in a few other countries. The DELPHI forecasts have been incorporated into the ensemble forecast from the \citet{CDC} and have been utilized in selecting the Phase III trial locations for the Johnson and Johnson COVID-19 vaccine. Historically, the DELPHI model has featured excellent predictive performance, matching the number of detected cases and deaths with high accuracy across the various waves of the pandemic.

In this paper, we integrate the (predictive) DELPHI model into a (prescriptive) optimization model for vaccine allocation. We first propose an extension of DELPHI, referred to as DELPHI--V, to capture the effects of vaccinations on the dynamics of the pandemic. The DELPHI--V model also disaggregates the dynamics of the pandemic at the subpopulation level to reflect disparities in mortality rates across age groups, which are critical drivers of vaccination strategies. We then formulate an optimization model, referred to as DELPHI--V--OPT, which optimizes the vaccine distribution strategy (that is, the deployment of mass vaccination sites at the strategic level, and the subsequent allocation of vaccines at the tactical level) to minimize the death toll of the pandemic. 
{\color{black}
Our focus on mass vaccination centers does not hinder the role that smaller vaccination sites (e.g., pharmacies) have been playing throughout the country to vaccinate the population. Ideally, our modeling approach would consider these various sites jointly. However, given the lack of publicly available information on the vaccines administered in smaller sites and the lack of coordination between the various vaccination sites, we leave this integration for future research.
}

From a technical standpoint, the DELPHI--V--OPT model relies on time discretization to embed the system of ordinary differential equations governing the DELPHI--V dynamics into an optimization model. The model is formulated as a bilinear (non-convex) optimization model, due to the SEIR dynamics at the core of DELPHI--V in which the number of new cases is driven by the number of susceptible and infected people. To solve it efficiently in realistic large-scale settings, we propose a coordinate descent algorithm. Starting from a baseline solution, the algorithm iterates, until convergence, between optimizing the vaccine distribution strategy (for given dynamics of the pandemic) and simulating the dynamics of the pandemic (for a given vaccine distribution strategy).

We implement the proposed model and algorithm using real-world data in the United States from the \cite{nyt:20}, the \cite{uscensus:01}, and the \cite{cdc:death_count}. We leverage the parameter estimates from the DELPHI model in each US state. One challenge, however, is that DELPHI estimates mortality rates in each state in each time period, while the \cite{cdc:death_count} reports mortality rates in each age bracket. To develop realistic and consistent estimates for mortality rates in each state, each age group and each time period, we formulate another bilinear optimization model that interpolates these two pieces of information, while ensuring consistency with broader demographic information.

Results suggest that the locations of vaccination sites can have a massive impact on the effectiveness of the vaccination campaign. As compared to several benchmarks based on demographic information (e.g., city and state population) and epidemiological information (e.g., case counts), our optimization approach increases the number of lives saved by the vaccines by 20\%, or 4,000 lives over a three-month period in the United States. These results underscore the necessity to consider both demographics and epidemiological dynamics when determining the locations of vaccination sites and subsequent vaccine allocation, which is achieved by the combination of our DELPHI--V epidemiological model and our optimization framework. In addition, the optimization approach can ensure equity between states and across vaccination sites, thus alleviating the death toll of the pandemic in some states without hurting others. Finally, these benefits are highly robust to misspecifications and fluctuations in the DELPHI parameters. Practically speaking, even though tactical decisions (e.g., vaccine allocation) need to be revised continuously in response to the latest information available throughout the vaccination campaign, strategic decisions (i.e., the location of vaccination sites) are highly robust to noise and uncertainty.

In summary, this paper makes three contributions. From a modeling standpoint, it formulates a novel optimization model for vaccine allocation, DELPHI--V--OPT, that integrates a state-of-the-art epidemiological model into an optimization model that supports vaccine distribution strategies, in order to mitigate the impact of the pandemic. From a computational standpoint, it develops a scalable coordinate descent algorithm, which converges effectively and in short runtimes. From a practical standpoint, it demonstrates that optimizing the locations of mass vaccination sites can curb the death toll of COVID-19 by a sizeable amount, thus highlighting the critical role of vaccine distribution besides vaccine design and vaccine production in combating the pandemic. Obviously, vaccine distribution involves broad political, economic and social considerations, which lie beyond the scope of this paper; yet, this paper can play a critical role to support ongoing mass vaccination efforts in order to mitigate the impact of the pandemic on public health.

\section{Literature review}
\label{literature_review}

Many pharmaceutical companies and academic institutions have explored different technologies toward a SARS-CoV-2 vaccine \citep{shin2020covid,florindo2020immune}. These span (i) inactivated or live-attenuated virus vaccines, which induce an immune response from weakened or killed pathogens (used by the Wuhan Institute of Biological Products, for instance); (ii) viral vector vaccines, which exploit non-replicating adenoviruses to deliver an antigenic element (used by Johnson and Johnson, for instance); (iii) subunit vaccines, which use a minimal structural component of a pathogen such as a protein (used by Clover Biopharmaceuticals, for instance); (iv) nucleic acid vaccines, which deliver DNA or mRNA of viral proteins (used by Pfizer and Moderna, for instance).

From an operational standpoint, a vast literature studies vaccine supply chains \citep[see][]{duijzer2018literature,lemmens2016review}. A first area involves optimizing vaccine composition \citep{wu2005optimization,kornish2008repeated,cho2010optimal,bandi2020optimizing}. A second area focuses on vaccine production to manage supply-side and demand-side uncertainty and mitigate incentive misalignments between manufacturers and end users \citep{chick2008supply,federgruen2009competition,arifouglu2012consumption}. Next, vaccine allocation optimizes the management of a vaccine stockpile \citep{sun2009selfish,mamani2013game}. Last, vaccine delivery optimizes inventory, distribution and dispensing operations \citep{jacobson2006analysis,aaby2006montgomery,dai2016contracting}. Most of this research focuses on predictable and repeatable epidemics, such as seasonal influenza. For less predictable epidemics, such as pandemic influenza, advance planning interventions include stockpiling \citep{jacobson2006stockpile} and anticipatory vaccination \citep{arinaminpathy2012impact}. Unfortunately, these approaches are not readily applicable to a new disease such as COVID-19.

Our paper deals with centralized vaccine allocation within a population. Early studies established the importance of partitioning the population into risk classes (e.g., age groups) to reflect the impact of an epidemic \citep{watson:72, elveback:76, longini:78}. \cite{emanuel2006should} propose a life-cycle model that prioritizes the most valuable subpopulations. Within a region, results suggest prioritizing at-risk populations \citep{patel2005finding,chowell2009adaptive} or active agents who can spread the disease fastest, such as school children \citep{dushoff2007vaccinating,basta2009strategies,medlock2009optimizing,lee2012modeling,matrajt2013optimal}. Across regions, results suggest that vaccines should be allocated to the most infected regions and to those affected the latest by the epidemic \citep{araz2012geographic,keeling2012optimal}.

Methodologically, most studies integrate SEIR or similar epidemiological models into simple optimization routines based on scenario analysis, enumeration, simulation, or simple heuristics \citep{uribe2011predictive,teytelman2013multiregional}. \cite{tanner2008finding} propose a chance-constrained optimization approach to ensure that the post-vaccination reproduction number is lower than one with high probability. \citet{yarmand2014optimal} formulate a two-stage stochastic programming model to first plan vaccine allocation and then distribute additional doses where the epidemic has not been contained. They model the dynamics of disease propagation by means of a stochastic SEIR model, and define scenarios using Monte Carlo simulation. In contrast, this paper directly embeds SEIR dynamics into an optimization model to support vaccine distribution.

Finally, this paper contributes to the fast-growing field of vaccine distribution in the midst of the COVID-19 pandemic. Recent and ongoing research spans vaccine production \citep{khamsi2020if}, equity in vaccine distribution \citep{muriel2021vaccine,bae2020challenges}, and public acceptance \citep{dror2020vaccine,coustasse2021covid}. In terms of vaccine distribution, \citet{rastegar2021inventory} propose a mixed-integer formulation to support influenza vaccine distribution during the COVID-19 pandemic. \citet{matrajt2020vaccine} study which populations to prioritize in a mass vaccination campaign, trading off vaccinating high-risk (older) age-groups vs. high-transmission (younger) age-groups in a given location. In contrast, this paper optimizes the distribution of vaccines across locations. This relates to \citet{grauer2020strategic}, who study the spatiotemporal distribution of vaccines, using an SEIR model to test various strategies based on demographic and epidemiological factors.

This paper expands this recent body of work in three major ways. First, we optimize vaccine allocation across regions and risk classes (e.g., age groups), based on data-driven estimates of infection and mortality rates. Second, we leverage a recent SEIR-inspired epidemiological model that captures dynamics specific to the COVID-19 pandemic, such as under-detection, governmental response, and declining mortality rates. Third, we propose a formal optimization approach and a coordinate descent algorithm to explicitly optimize vaccine distribution strategies, as opposed to relying on enumeration, simulation or simplified heuristics.

\section{Model formulation}
\label{model_formulation}

Our model optimizes vaccine distribution strategy. In the US context, this primarily involves the location of mass vaccination sites. However, optimizing these decisions requires to account for subsequent vaccine allocation across the population, in order to further optimize and evaluate the effects of the vaccination campaign. Therefore, we refer to as \emph{vaccine distribution strategy} the set of three decisions: (i) the location of mass vaccination sites, (ii) the allocation of vaccines across vaccination sites, and (iii) the allocation of vaccines within each sub-population.

We capture the dynamics of the pandemic by means of an epidemiological model, called DELPHI, which forecasts the number of detected cases, hospitalizations and deaths in each US state \citep{li:20}.\footnote{DELPHI is also applied to each country and to other provinces, but this paper focuses on US states.} We review it briefly, and augment it to capture the effects of vaccinations---we refer to this model as DELPHI--V. We then embed the DELPHI--V model into a mathematical programming model to optimize vaccine allocation, referred to as DELPHI--V--OPT.

\subsection{DELPHI: Forecasting the dynamics of the COVID-19 pandemic}

DELPHI is a compartmental epidemiological model, which extends the widely used SEIR model to account for specificities of the COVID-19 pandemic. The model is governed by a system of ordinary differential equations (ODEs) across 11 states: susceptible ($S$), exposed ($E$), infectious ($I$), undetected cases who will recover ($U^R$) or die ($U^D$), hospitalized cases who will recover ($H^R$) or die ($H^D$), quarantined cases who will recover ($Q^R$) or die ($Q^D$), recovered ($R$) and dead ($D$).

DELPHI differs from most other COVID-19 forecasting models \citep[see, e.g.][]{kissler:20,notredame,rodriguez2020deepcovid} by capturing three key elements of the pandemic: 
\begin{itemize}
    \item \textbf{Under-detection}: Many cases remain undetected due to limited testing, asymptomatic carriers, and detection errors. Ignoring them would underestimate the scale of the pandemic. The DELPHI model captures them through the $U^R$ and $U^D$ states.
    \item \textbf{Governmental and societal response}: Social distancing policies limit the spread of the virus. Ignoring them would overestimate the scale of the pandemic. However, if restrictions are lifted prematurely, a resurgence may occur. We define a governmental and societal response function $\gamma(t)$, which modulates the infection rate and is parameterized as follows: 
    \begin{align}
        \gamma(t)&=1 + \frac{2}{\pi}\arctan\left(\frac{-(t-t_\text{int})}{\omega}\right) + c\exp\left(-\frac{(t-t_\text{jump})^2}{2\sigma^2}\right). \label{eq:gamma}
    \end{align}
    This parameterization defines four phases (Figure~\ref{fig:response_function}). In Phase I, most activities continue normally. In Phase II, the infection rate declines sharply as policies get implemented. The parameters $t_{\text{int}}$ and $\omega$ can be interpreted as the start time and strength of this response. In Phase III, the decline reaches saturation. The epidemic then experiences a resurgence of magnitude $c$ in Phase IV, due to relaxations in governmental and social restrictions. This is counteracted at time $t_{\text{jump}}$, when restrictions are re-implemented, with $\sigma$ controlling the duration of this second wave.
\begin{figure*}[ht!]
    \centering
    \includegraphics[width=.8\textwidth]{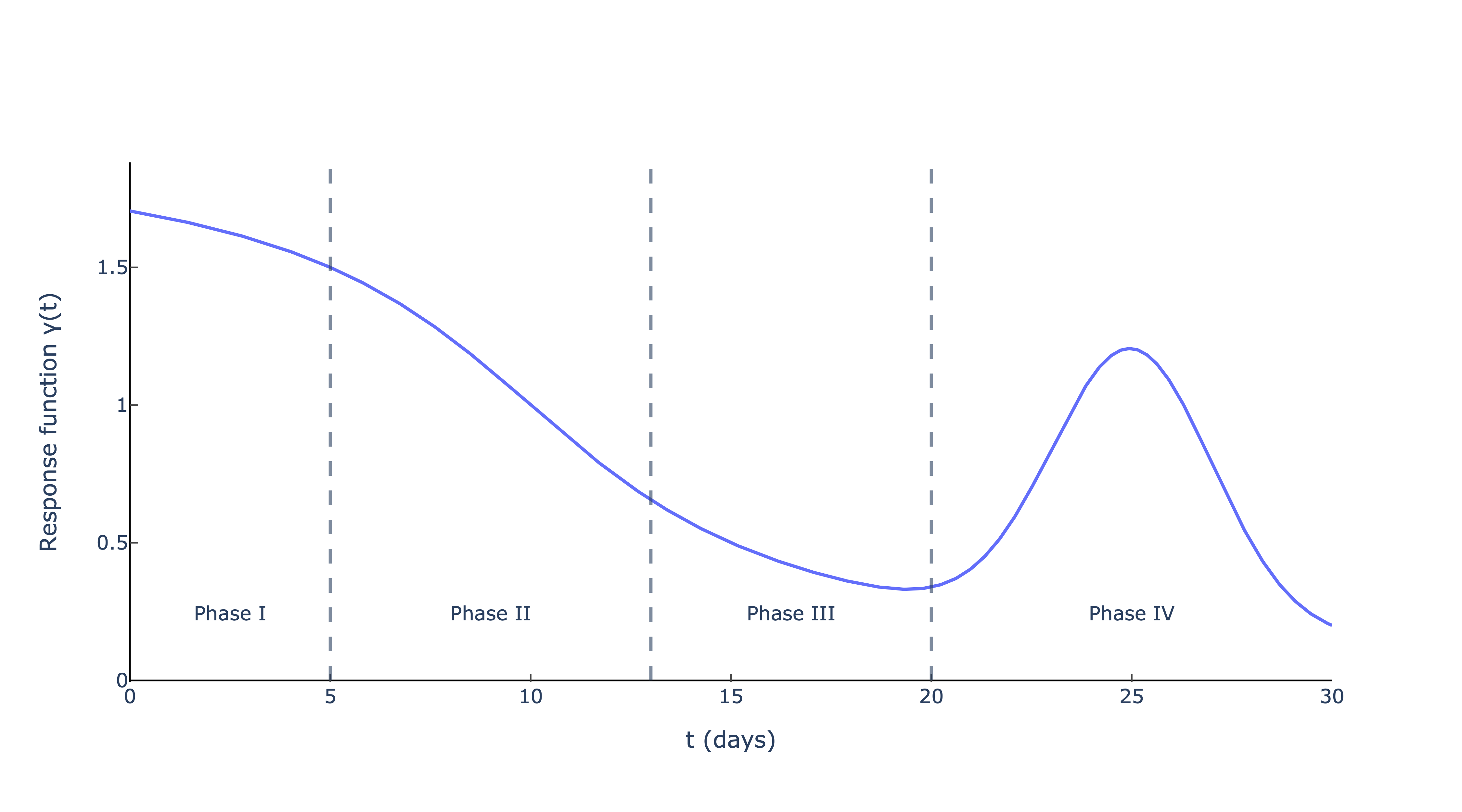}
    \caption{Governmental and societal response function $\gamma(t)$ ($\omega=5$, $t_{\text{int}}=10$, $c=1$, $t_{\text{jump}}=25$ and $\sigma=2$).}
    \label{fig:response_function}
\end{figure*}
    \item \textbf{Declining mortality rates}: The mortality rate of COVID-19 has been declining through the pandemic, due to a better detection of mild cases, enhanced care for COVID-19 patients, and other factors. We model the mortality rate as a monotonically decreasing function of time:
    \begin{align}
        m(t) &= \left(m_0 - m_{\min}\right)\left(1 + \frac{2}{\pi}\arctan\left(-r_m t\right)\right) + m_{\min}, \label{eq:mortality}
    \end{align}
    where $m_0$ is the initial mortality rate, $m_{\min}$ is the minimum mortality rate and $r_m$ is a decay rate. 
\end{itemize}

Ultimately, DELPHI involves 16 parameters that define the transition rates between the 11 states. We calibrate seven of them from a database on clinical outcomes \citep{bertsimas:20}. Using non-linear optimization, we estimate the other 9 parameters from historical data on the number of cases and deaths in each region. We refer to \citet{li:20} for details.

\begin{figure}[ht]
\begin{tabular}{@{}p{\halfwidth}p{\halfwidth}@{}}
\raisebox{-\height}{ \includegraphics[width=0.48\textwidth]{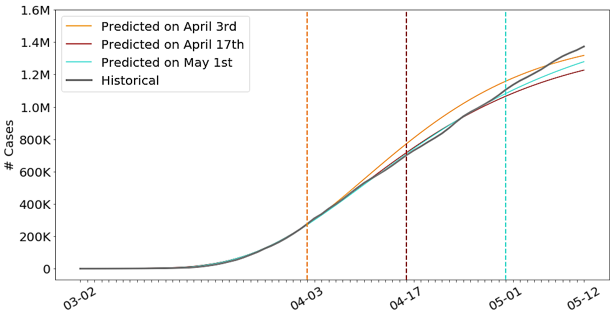}}&
    \raisebox{-\height}{\includegraphics[width=0.48\textwidth]{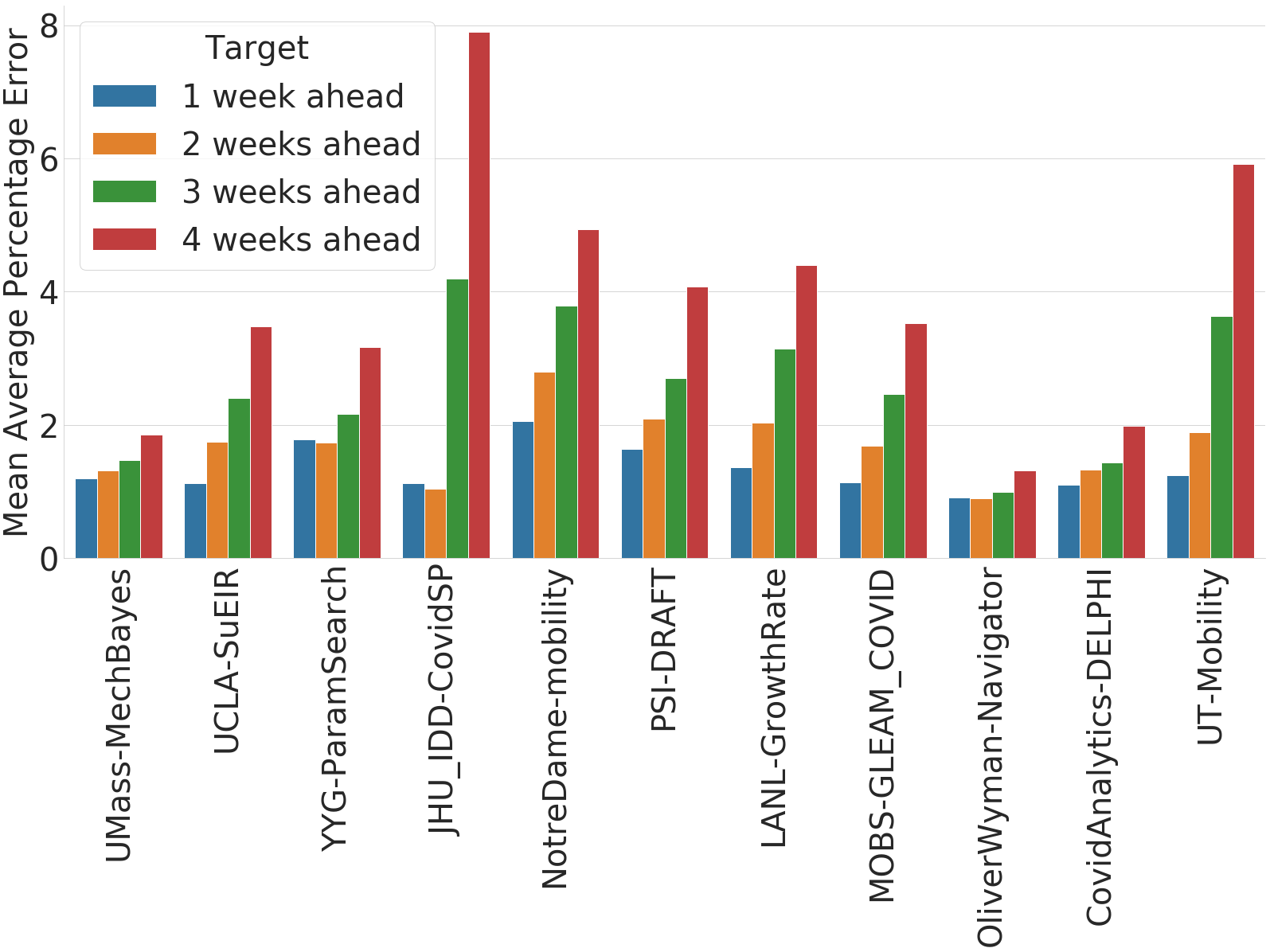}}\\
    	\subcaption{\label{subfig:DELPHI1}US First-wave predictions} & \subcaption{\label{subfig:DELPHI2}Second-wave performance} 
\end{tabular}
	\vspace{6 pt}
	\caption{Historical performance of the DELPHI predictions in the United States.}
	\label{fig:DELPHI}
\end{figure}

Since its inception in March 2020, DELPHI has been extensively tested and validated against real-world data. Figure~\ref{fig:DELPHI} reports the historical performance of the model in the United States, during the first wave in the Spring of 2020 and the second wave in the Fall of 2020. As the figure shows, the model has been predicting the magnitude of the pandemic with high accuracy up to one month in advance; for instance, as early as April 3, 2020, the model was predicting 1.2--1.4 million cases in the United States by early May, a prediction that became quite accurate a month later (Figure~\ref{subfig:DELPHI1}). Obviously, subsequent forecasts, by leveraging more up-to-date information, were able to refine these estimates. As a result, the DELPHI model was incorporated into the ensemble forecast from the \citet{CDC}. During the second wave of the pandemic, DELPHI continued to exhibit strong predictive performance, with a mean average percentage error among the lowest of the CDC ensemble forecast (Figure~\ref{subfig:DELPHI2}).

\subsection{Predictive DELPHI--V: Capturing the effects of vaccination}
\label{delphi_with_vaccinations}

We now augment the DELPHI model to capture two key aspects of vaccinations:
\begin{enumerate}
\item \emph{Disparate impacts of the disease across risk classes}. Age is one of the primary drivers of mortality \citep{guan2020clinical,goyal2020clinical, petrilli2020factors}. The \cite{cdc:death_mult} reports that the mortality rate among Americans aged 70 and over is two orders of magnitude greater than for those aged 30 and under. We partition the population into \emph{risk classes}, defined as homogeneous groups with comparable health characteristics. We consider age-based risk classes in our experiments, but other categorizations could be used (e.g., based on comorbidities). Accordingly, we replicate the 11 model states for each risk class.
\item \emph{Impact of vaccinations on the dynamics of the pandemic}. A fraction of vaccinated people will be immune to the disease (based on the vaccine's effectiveness). Clinical trials suggest that early-approved vaccines prevent mortality but not necessarily infections. Therefore, we assume conservatively that all vaccinated people can still transmit the disease. We relax this assumption later on, to show the robustness of our results when a fraction of vaccinated people become fully immune to the disease. We create four new model states: susceptible and vaccinated ($S'$), exposed and vaccinated ($E'$), infected and vaccinated ($I'$), and immune ($M$).
\end{enumerate}

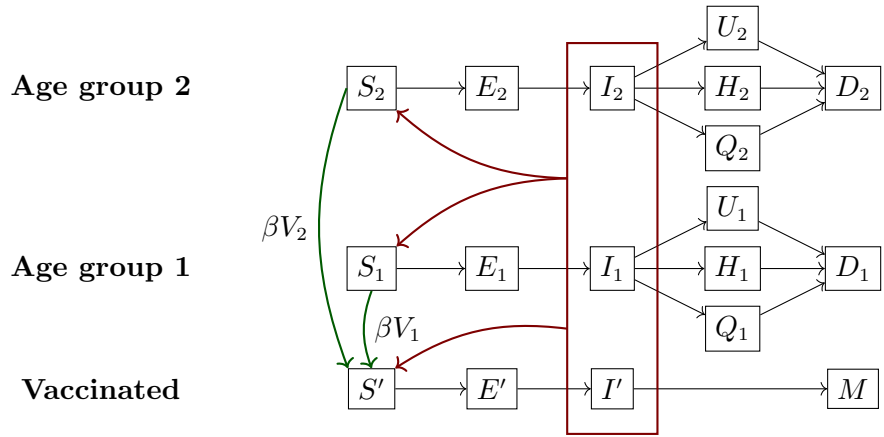
\begin{figure*}[h!]
\centering
\begin{tikzpicture}[scale=0.4, every node/.style={scale=1}]
    \node[] at (-9,0) {\textbf{Vaccinated}};
    \node[] at (-9,4) {\textbf{Age group 1}};
    \node[] at (-9,10) {\textbf{Age group 2}};
    
    \node[draw=black] (S1) at (0, 4){$S_1$};
    \node[draw=black] (E1) at (4, 4){$E_1$};
    \node[draw=black] (I1) at (8, 4) {$I_1$};
    
    \node[draw=black] (UD1) at (12, 6) {$U_1$};
    \node[draw=black] (HD1) at (12, 4) {$H_1$};
    \node[draw=black] (QD1) at (12, 2) {$Q_1$};

    \node[draw=black] (D1) at (16,4) {$D_1$};

    \draw[->] (S1)--(E1);
    \draw[->] (E1)--(I1);
    \draw[->, draw=black] (I1)--(HD1);
    \draw[->, draw=black] (I1)--(QD1);
    \draw[->, draw=black] (I1)--(UD1);

    \draw[->, draw=black] (UD1)--(D1);
    \draw[->, draw=black] (HD1)--(D1);
    \draw[->, draw=black] (QD1)--(D1);
    
    \node[draw=black] (S2) at (0, 10){$S_2$};
    \node[draw=black] (E2) at (4, 10){$E_2$};
    \node[draw=black] (I2) at (8, 10) {$I_2$};
    
    \node[draw=black] (UD2) at (12, 12) {$U_2$};
    \node[draw=black] (HD2) at (12, 10) {$H_2$};
    \node[draw=black] (QD2) at (12, 8) {$Q_2$};

    \node[draw=black] (D2) at (16,10) {$D_2$};
    
    \draw[->] (S2)--(E2);
    \draw[->] (E2)--(I2);
    \draw[->, draw=black] (I2)--(HD2);
    \draw[->, draw=black] (I2)--(QD2);
    \draw[->, draw=black] (I2)--(UD2);

    \draw[->, draw=black] (UD2)--(D2);
    \draw[->, draw=black] (HD2)--(D2);
    \draw[->, draw=black] (QD2)--(D2);
    
    \node[draw=black] (SV) at (0, 0){$S'$};
    \node[draw=black] (EV) at (4, 0){$E'$};
    \node[draw=black] (IV) at (8, 0) {$I'$};
    
    \node[draw=black] (Immune) at (16,0) {$M$};
    
    \draw[->] (SV)--(EV);
    \draw[->] (EV)--(IV);
    \draw[->] (IV)--(Immune);
    \draw[->, draw=mygreen, thick] (S1.south) to [bend right=20] node[right,midway]{$\beta V_1$} (SV.north);
    \draw[->, draw=mygreen, thick] (S2.west)  to [bend right=20] node[left,midway]{$\beta V_2$} (SV.north west);
    
    \draw[draw=myred, thick] (6.5,-1.5) rectangle (9.5,11.5);
    \draw[->, draw=myred, thick] (6.5,7) to [bend left=20] (S2.south east);
    \draw[->, draw=myred, thick] (6.5,7) to [bend right=20] (S1.north east);
    \draw[->, draw=myred, thick] (6.5,2) to [bend right=20] (SV.north east);
    \end{tikzpicture}
\caption{Simplified flow diagram of the DELPHI--V model.}
\label{fig:DELPHI_with_vaccinations}
\end{figure*}

Figure~\ref{fig:DELPHI_with_vaccinations} shows a simplified flow diagram of the DELPHI--V model, with two risk classes (indexed by $k=1,2$ and indicated via subscripts). For expositional purposes, we omit dependencies on the region, since the DELPHI--V model is fitted in each region independently. In the remainder of this paper, we also ignore the recovery states, since they do not impact the death-minimization optimization model. Accordingly, we denote the states of undetected, hospitalized and quarantined people who will die from the disease by $U$, $H$ and $Q$ (as opposed to $U^D$, $H^D$ and $Q^D$).

For simplicity, we make three assumptions. First, the effects of vaccines are instantaneous (relaxing this assumption, although straightforward, would merely induce a time lag into the system, without significantly impacting the vaccine distribution strategy). Second, the vaccine has no effect when it fails to immunize the patient (i.e., no partial benefit and no side effect). Third, we consider single-dose vaccines. {\color{black}In reality, vaccines can require a single dose or two doses. Double-dose vaccines could be modeled by adding another state of one-dosed patients between $S_k$ and $S'$ (similar to the construction in \cite{mak2021managing}). This modeling extension would raise new questions surrounding the likelihood of one-dosed patients to contract, transmit and die from the disease---all of which involve significant uncertainties in the absence of relevant data. In addition, given the heterogeneity of vaccines currently available, this extended model extension would end up determining which states get which \emph{type} of vaccines. These decisions, however, are mainly driven by supply chain considerations rather than epidemiological considerations. Therefore, we focus on single-dose vaccines in this paper, and lead the integration of double-dose vaccines for future research.

Given these assumptions, the model captures the effects of vaccinations as follows.} Let $V_k(t)$ denote the population mass from risk class $k\in\mathcal{K}$ that gets vaccinated at time $t\in\mathcal{T}$, and let $\beta\in(0, 1]$ denote the vaccine's effectiveness. A mass $\beta V_k(t)$ of people transitions from the susceptible state $S_k$ to the state $S'$, and the remaining mass $(1-\beta)V_k(t)$ remains in the susceptible state. People in the $S'$ state can become exposed and infected, but then become immune to the disease (as opposed to having a positive probability of dying from it). Note that infections are driven by the total mass of infected people, across all risk classes and vaccinated people. All other transitions shown in Figure~\ref{fig:DELPHI_with_vaccinations} are consistent with the DELPHI model.

The DELPHI--V model is governed by the following ODE system:
\begin{align}
    \label{eq:susceptible_ode}
    \frac{\text{d}S_k}{\text{d}t} &= -\beta V_k(t)-\alpha\gamma(t)\left(S_k(t)-\beta V_k(t)\right)\left(\sum_{l=1}^K I_l(t)+I'(t)\right) \\
    \label{eq:susceptibleV_ode}
    \frac{\text{d}S'}{\text{d}t} &= +\beta \sum_{\kappa=1}^KV_\kappa(t)-\alpha\gamma(t)\left(S'(t)+\beta \sum_{\kappa=1}^KV_\kappa(t)\right)\left(\sum_{l=1}^K I_l(t)+I'(t)\right) \\
    \label{eq:exposed_ode}
    \frac{\text{d}E_k}{\text{d}t} &= \alpha\gamma(t)\left(S_k(t)-\beta V_k(t)\right)\left(\sum_{l=1}^K I_l(t)+I_V(t)\right) - r^I E_k(t) \\
    \label{eq:exposedV_ode}
    \frac{\text{d}E'}{\text{d}t} &= \alpha\gamma(t)\left(S'(t)+\beta \sum_{\kappa=1}^KV_\kappa(t)\right)\left(\sum_{l=1}^K I_l(t)+I'(t)\right) - r^I E'(t) \\
    \label{eq:infectious_ode}
    \frac{\text{d}I_k}{\text{d}t} &= r^I E_k(t) - r^dI_k(t) \\
    \label{eq:infectiousV_ode}
    \frac{\text{d}I'}{\text{d}t} &= r^I E'(t) - r^dI'(t) \\
    \label{eq:undetected_dying_ode}
    \frac{\text{d}U_k}{\text{d}t} &= r_k^U(t) I_k(t) - r^D U_k(t) \\
    \label{eq:hospitalized_dying_ode}
    \frac{\text{d}H_k}{\text{d}t} &= r_k^H(t) I_k(t) - r^D H_k(t) \\
    \label{eq:quarantined_dying_ode}
    \frac{\text{d}Q_k}{\text{d}t} &= r_k^Q(t) I_k(t) - r^D Q_k(t) \\
    \label{eq:deceased_ode}
    \frac{\text{d}D_k}{\text{d}t} &= r^D \left(U_k(t) + H_k(t) + Q_k(t)\right) \\
    \label{eq:immune_ode}
    \frac{\text{d}M}{\text{d}t} &= r^dI'(t),
\end{align}
where:
\begin{itemize}
    \item[--] $\alpha$ is the nominal infection rate;
    \item[--] $\gamma(t):\mathbb{R}_+\rightarrow\mathbb{R}_+$ is the governmental and societal response function (Figure~\ref{fig:response_function});
    \item[--] $r^I$, $r^d$, $r^D$, are the progression rate, the detection rate, and the death rate;
    \item[--] $r_k^U(t)$, $r_k^H(t)$, and $r_k^Q(t)$ capture the detection, hospitalization and death rates, accounting for the probabilities of detection and hospitalization and the mortality rate (Equation~\eqref{eq:mortality}). Their dependency on $t$ and $k$ reflect disparities over time and across risk classes.
\end{itemize}

As noted earlier, the dynamics of exposure and infection depend on the total number of infected people (across risk classes and vaccinated/non-vaccinated people), as opposed to the number of infected people in a given risk class. DELPHI--V captures these interdependencies---indicated by the red rectangle in Figure~\ref{fig:DELPHI_with_vaccinations} and the terms $\sum_{l=1}^K I_l(t)+I'(t)$ in Equations~\eqref{eq:susceptible_ode}--\eqref{eq:exposed_ode}.

Given initial conditions, the ODE equations uniquely determine the evolution of this system over time---for a given vaccine allocation reflected in the variable $\mathbf{V}$. Next, we optimize the vaccine distribution strategy to minimize the overall impact of the pandemic---estimated by DELPHI--V.

\subsection{Prescriptive DELPHI--V--OPT: Optimizing the vaccine distribution strategy}
\label{prescriptive_delphi}

The DELPHI--V--OPT model takes as inputs epidemiological information (estimated from the DELPHI--V model), information on the vaccine (including vaccine effectiveness and vaccine budget), and demographic information in the United States (e.g., major cities, distance across counties, population per county). It optimizes the vaccine distribution strategy, including the location of mass vaccination sites and the subsequent allocation of vaccines. It is formulated as a tri-objective model, to minimize (i) the death toll of the pandemic, (ii) the number of exposed people in the termination period, and (iii) the distance between vaccination sites and population centers. The main public health objective is obviously death minimization, so the first objective component is heavily prioritized. However, just considering the number of deaths could result in a waste of vaccines near the end of the planning horizon, as individuals infected in the final periods would not have time to flow to the death state in the epidemiological model. Therefore, the second component of the objective minimizes the number of infections. The last component minimizes geographic disparities. In addition, the model incorporates other equity consideration by means of fairness constraints.

We proceed by time discretization to formulate the optimization model and retain tractability. This reduces to solving the system of ODE equations given in Equations~\eqref{eq:susceptible_ode}--\eqref{eq:immune_ode} by a forward difference scheme. We denote by $\Delta t$ the discretization unit (e.g., 1 day).

Formally, we define the following sets, input parameters, and decision variables.
\paragraph{Sets}
{\setlength{\jot}{-3pt}\setlength{\abovedisplayskip}{0pt}\setlength{\belowdisplayskip}{0pt}\setlength{\abovedisplayshortskip}{0pt}\setlength{\belowdisplayshortskip}{0pt}
\begin{align*}
\mathcal{L}							&=\text{set of population centers in the United States (e.g., counties), }\lbrace 1,\cdots,L\rbrace					\\
\mathcal{M}							&=\text{set of regions in the United States (e.g., 50 states plus the District of Columbia), }\lbrace 1,\cdots,m\rbrace	\\
\mathcal{N}							&=\text{set of candidate vaccination sites, }\lbrace 1,\cdots,n\rbrace											    \\
\mathcal{N}_j                       &=\text{subset of candidate vaccination sites located in state $j\in\mathcal{M}$}                                   \\
\mathcal{N}_l                       &=\text{subset of candidate vaccination sites located in the same state as county $l\in\mathcal{L}$}                \\
\mathcal{K}							&=\text{set of risk classes in the population (e.g., age groups), }\lbrace 1,\cdots,K\rbrace    					\\
\mathcal{T}							&=\text{set of discretized time periods (e.g., days), }\lbrace 1,\cdots,T\rbrace										
\end{align*}}
\vspace{-6 pt}
\paragraph{Parameters}
{\setlength{\jot}{-3pt}\setlength{\abovedisplayskip}{0pt}\setlength{\belowdisplayskip}{0pt}\setlength{\abovedisplayshortskip}{0pt}\setlength{\belowdisplayshortskip}{0pt}
\begin{align*}
Pop_l                               &=\text{total population in center $l\in\mathcal{L}$}                                                                                            \\
\Delta_{li}							&=\text{distance from population center $l\in\mathcal{L}$ to candidate vaccination site $i\in\mathcal{N}$}                                                  \\
N     							    &=\text{number of vaccination sites to be deployed across the country}                                                                                      \\
B_t     							&=\text{number of vaccines available at time $t\in\mathcal{T}$}                                                                                             \\
\beta     							&=\text{effectiveness of vaccines}                                                                                                                          \\
r^I                                 &=\text{progression rate of the disease (DELPHI parameter)}                                                                                                                    \\
r^d                                 &=\text{detection rate of the disease (DELPHI parameter)}                                                                                                                      \\
r^D                                 &=\text{death rate of the disease (DELPHI parameter)}                                                                                                                          \\
r_{jkt}^U                           &=\text{transition rate from $I$ to $U$, capturing mortality rate} \\
                                    &\quad\text{in region $j\in\mathcal{M}$ at time $t\in\mathcal{T}$ for risk class $k\in\mathcal{K}$ (DELPHI parameter)}   \\
r_{jkt}^H                           &=\text{transition rate from $I$ to $H$, capturing mortality rate} \\
                                    &\quad\text{in region $j\in\mathcal{M}$ at time $t\in\mathcal{T}$ for risk class $k\in\mathcal{K}$ (DELPHI parameter)}   \\
r_{jkt}^Q                           &=\text{transition rate from $I$ to $Q$, capturing mortality rate} \\
                                    &\quad\text{in region $j\in\mathcal{M}$ at time $t\in\mathcal{T}$ for risk class $k\in\mathcal{K}$ (DELPHI parameter)}   \\
r_{jkt}^H                           &=\text{transition rate from $I$ to $H$, capturing mortality rate} \\
\alpha_j                            &=\text{nominal infection rate in region $j\in\mathcal{M}$ (DELPHI parameter)}                                                                                                 \\
\gamma_{jt}                         &=\text{governmental and societal response in region $j\in\mathcal{M}$ at time $t\in\mathcal{T}$ (DELPHI parameter)}
\end{align*}}

Note that the parameters $r^U$, $r^H$ and $r^Q$ are defined for each region, risk class and time period, reflecting underlying variations in mortality rates. In contrast, the parameters $r^I$, $r^d$ and $r^D$ are treated as uniform characteristics of the disease. In reality, these parameters may vary across risk classes; for instance, the serological estimates from the \citet{CDC} suggest different prevalence of the disease across age groups. We test this hypothesis in our experiments, to verify the robustness of our results to the uniform infection rate assumption.

We also assume a single vaccine effectiveness value $\beta$. In theory, vaccine effectiveness might also vary across risk classes. More importantly, there are now several vaccines available, each with different clinical characteristics. Ideally, we could introduce an additional set to capture vaccine heterogeneity, and let the vaccine effectiveness vary across vaccine types. This approach however, may be somewhat impractical in practice, as it may be difficult to strategically allocate different vaccines to different populations based on vaccine effectiveness. For equity, we therefore assume conservatively that the mix of vaccines remains identical across vaccination sites. Under this restriction, the mix of vaccines can be reduced to a representative vaccine with average effectiveness.

\paragraph{Primary decision variables}

\begin{align*}
    x_i=&\quad\begin{cases}1&\text{if vaccination site $i\in\mathcal{N}$ is selected}\\0&\text{otherwise}\\\end{cases}\\
    C_{it}:&\quad\text{number of vaccines distributed to site $i\in\mathcal{N}$ at time $t\in\mathcal{T}$}\\
    W_{li}=&\quad\begin{cases}1&\text{if population center $l\in\mathcal{L}$ is assigned to vaccination site $i\in\mathcal{N}$}\\0&\text{otherwise}\\\end{cases}\\
    \bar{S}_{jkt}:&\quad\text{number of eligible people to region $j\in\mathcal{M}$ in risk class $k\in\mathcal{K}$ at time $t\in\mathcal{T}$}\\
    V_{jkt}:&\quad\text{number of vaccines allocated to region $j\in\mathcal{M}$ in risk class $k\in\mathcal{K}$ at time $t\in\mathcal{T}$}
\end{align*}

To track the impact of vaccine allocation on the resulting dynamics of the pandemic, we create indirect variables, corresponding to all the states in the DELPHI--V model shown in Figure~\ref{fig:DELPHI_with_vaccinations}.

The vaccine distribution problem, referred to as $(\mathcal{P})$, is then formulated in Equations~\eqref{eq:objective}--\eqref{eq:domain_constraint}.

{\small
\begin{align}
    \label{eq:objective}
    \min ~&\  \sum_{j=1}^{M}\sum_{k=1}^K \left(D_{jkT} + H_{jkT} + Q_{jkT}\right)+\lambda_E\sum_{j=1}^{M}\sum_{k=1}^KE_{jkT}+\lambda_D\sum_{l=1}^L\sum_{i=1}^nPop_l\Delta_{li}W_{li} \\
    \label{eq:total_cap}
    \text{s.t.}~&\  \sum_{i=1}^nx_i=N  \\
    \label{eq:one_per_state}
    &\  \sum_{i\in\mathcal{N}_j}x_i\geq 1, && \forall j\in\mathcal{M}  \\
    \label{eq:assign_consistency}
    & \  W_{li}\leq x_i, && \forall l\in\mathcal{L},\  i\in\mathcal{N}   \\
    \label{eq:assign_all}
    & \  \sum_{i\in \mathcal{N}_l}W_{li}=1, && \forall l\in\mathcal{L}   \\
    \label{eq:budget}
    & \  \sum_{i=1}^n C_{it}\leq B_t, && \forall t\in\mathcal{T}   \\
    \label{eq:dist_consistency}
    & \  C_{it}\leq B_tx_i, && \forall i\in\mathcal{N},\  t\in\mathcal{T}   \\
    \label{eq:alloc_consistency}
    & \  \sum_{k=1}^K V_{jkt} \leq \sum_{i \in \mathcal{N}_j} C_{it}, && \forall j\in\mathcal{M},\  t\in\mathcal{T}   \\
    \label{eq:eligibility_constraint}
    &\  \bar{S}_{j,k,t+1} \leq \bar{S}_{jkt} - (1-\beta) V_{jkt} - (S_{jkt}-S_{j,k,t+1}), && \forall j \in \mathcal{M}, k \in \mathcal{K}, t \in \mathcal{T} \\
    \label{eq:eligibility2_constraint}
    &\  V_{jkt} \leq \bar{S}_{jkt}, && \forall j \in \mathcal{M}, k \in \mathcal{K}, t \in \mathcal{T} \\
    \label{eq:smoothness_constraint}
    &\ \left|C_{i,t+1}-C_{it}\right| \leq \theta_SC_{it}, && \forall i \in \mathcal{N}, t \in \mathcal{T} \\
    \label{eq:fairness_sites}
    & \  \frac{\sum_{l\in\mathcal{L}_j}Pop_l}{\sum_{j'=1}^m\sum_{l\in\mathcal{L}_{j'}}Pop_l}N - \Theta_L \leq \sum_{i\in\mathcal{N}_j}x_i \leq \frac{\sum_{l\in\mathcal{L}_j}Pop_l}{\sum_{j'=1}^m\sum_{l\in\mathcal{L}_{j'}}Pop_l}N + \Theta_L, && \forall j\in\mathcal{M}   \\
    \label{eq:fairness_dist}
    & \  \frac{B_t}{N(1+\theta_V)}x_i \leq C_{it} \leq \frac{B_t}{N} (1 + \theta_V)x_i, && \forall i\in\mathcal{N},\  t\in\mathcal{T}   \\
    \label{eq:fairness_vaccines}
    & \  \sum_{i \in \mathcal{N}_j} C_{it} \leq \left(\frac{\sum_{l\in\mathcal{L}_j}Pop_l}{\sum_{j'=1}^m\sum_{l\in\mathcal{L}_{j'}}Pop_l} + \theta_P\right) B_t, && \forall j\in\mathcal{M},\  t\in\mathcal{T}   \\
    \label{eq:susceptible_constraint}
    &\  S_{j,k,t+1} \geq S_{jkt} - \beta V_{jkt} - \alpha_j\gamma{jt}\left(S_{jkt}-\beta V_{jkt}\right)\left(\sum_{\kappa=1}^K I_{j\kappa t}+I'_{jt}\right)\Delta t, && \forall j \in \mathcal{M}, k \in \mathcal{K}, t \in \mathcal{T} \\
    \label{eq:susceptibleV_constraint}
    &\  S'_{j,t+1} \geq S'_{jt} + \beta \sum_{\kappa=1}^KV_{j\kappa t} - \alpha_j\gamma_{jt}\left(S'_{jt} + \beta \sum_{\kappa=1}^KV_{j\kappa t}\right)\left(\sum_{\kappa=1}^K I_{j\kappa t}+I'_{jt}\right)\Delta t, && \forall j \in \mathcal{M}, t \in \mathcal{T} \\
    \label{eq:exposed_constraint}
    &\  E_{j,k,t+1} \geq E_{jkt} +\left( \alpha_j\gamma_{jt}\left(S_{jkt}-\beta V_{jkt}\right)\left(\sum_{\kappa=1}^K I_{j\kappa t}+I'_{jt}\right) - r^I E_{jkt}\right)\Delta t, && \forall j \in \mathcal{M},\  k \in \mathcal{K},\  t \in \mathcal{T}  \\
    \label{eq:exposedV_constraint}
    &\  E'_{j,t+1} \geq E'_{jt} + \left( \alpha_j\gamma_{jt}\left(S'_{jt} + \beta \sum_{\kappa=1}^KV_{j\kappa t}\right)\left(\sum_{\kappa=1}^K I_{j\kappa t}+I'_{jt}\right) - r^I E'_{jt}\right)\Delta t, && \forall j \in \mathcal{M}, t \in \mathcal{T} \\
    \label{eq:infectious_constraint}
    &\  I_{j,k,t+1} \geq I_{jkt} + \left(r^I E_{jkt} - r^dI_{jkt}\right)\Delta t, && \forall j \in \mathcal{M},\  k \in \mathcal{K},\  t \in \mathcal{T} \\
    \label{eq:infectiousV_constraint}
    &\  I'_{j,t+1} \geq I'_{jt} + \left(r^I E'_{jt} - r^dI'_{jt}\right)\Delta t, && \forall j \in \mathcal{M},\  t \in \mathcal{T} \\
    \label{eq:undetected_constraint}
    &\  U_{j,k,t+1} \geq U_{jkt} + \left(r^U_{jkt} I_{jkt} - r^DU_{jkt}\right)\Delta t,  && \forall j \in \mathcal{M},\  k \in \mathcal{K},\  t \in \mathcal{T} \\
    \label{eq:hospitalized_constraint}
    &\  H_{j,k,t+1} \geq H_{jkt} + \left(r^H_{jkt} I_{jkt} - r^D H_{jkt}\right)\Delta t, && \forall j \in \mathcal{M},\  k \in \mathcal{K},\  t \in \mathcal{T} \\
    \label{eq:quarantined_constraint}
    &\  Q_{j,k,t+1} \geq Q_{jkt} + \left(r^Q_{jkt} I_{jkt} - r^D Q_{jkt}\right)\Delta t, && \forall j \in \mathcal{M},\  k \in \mathcal{K},\  t \in \mathcal{T} \\
    \label{eq:deceased_constraint}
    &\  D_{j,k,t+1} = D_{jkt} + r^D\left(U_{jkt} + H_{jkt} + Q_{jkt}\right)\Delta t, && \forall j \in \mathcal{M},\ k \in \mathcal{K},\ t \in \mathcal{T} \\
    \label{eq:immune_constraint}
    &\  M_{j,t+1} = M_{jkt} + r^dI'_{jt}, &&\forall j \in \mathcal{M},\  t \in \mathcal{T} \\
    \label{eq:domain_constraint}
    &\  \mathbf{x,W}\text{ binary},\ \mathbf{C}, \mathbf{V}, \mathbf{S}, \mathbf{S'}, \mathbf{\bar{S}}, \mathbf{E}, \mathbf{E'}, \mathbf{I}, \mathbf{I'}, \mathbf{U}, \mathbf{H}, \mathbf{Q}, \mathbf{D}, \mathbf{M} \geq \mathbf{0}.
\end{align}
}
Equation~\eqref{eq:objective} formalizes the three objectives of the model. The first term corresponds to our primary objective of minimizing the number of deaths over the planning horizon, across all regions and risk classes. This number includes people in the absorbing state $D$, as well as the transient states $H$ and $Q$ (we ignore undetected deaths). The next terms minimize, as lower-priority objectives, the number of exposed people at the end of the horizon and the distance to the vaccination sites. The hyperparameters $\lambda_E$ and $\lambda_D$ are set to small values to prioritize the death-minimization objective.

Next, the constraints capture practical considerations surrounding vaccine distribution:
\begin{itemize}
    \item[--] \textbf{Number of vaccination sites}: We impose a total budget of $N$ vaccination sites (Equation~\eqref{eq:total_cap}). For obvious reasons, there needs to be at least one site in every state (Equation~\eqref{eq:one_per_state}). We consider $N=100$ in our experiments, which leaves flexibility to strategically deploy 49 sites.
    \item[--] \textbf{Assignment}: Equation~\eqref{eq:assign_consistency} ensures that people get assigned to vaccination sites that have been selected. Equation~\eqref{eq:assign_all} assigns each population center to exactly one site, in the same state. These assignment constraints are used to compute the distance term in the objective function.
    \item[--] \textbf{Inter-regional vaccine capacity}: Due to restrictions in vaccine manufacturing and distribution networks, a limited number of vaccines can be allocated in each time period. Equation~\eqref{eq:budget} ensures that the total number of vaccines allocated lies within the available budget in each period.
    \item[--] \textbf{Consistency}: Equation~\eqref{eq:dist_consistency} ensures that vaccines only get distributed to selected sites. Similarly, Equation~\eqref{eq:alloc_consistency} ensures that the number of people vaccinated in each state (across risk classes) does not exceed the number of vaccines allocated that state. This constraint involves two assumptions. A first, conservative assumption is that people can only get vaccinated in the state that they live in, which is required in practice for traceability purposes. Another, optimistic assumption is that the vaccine allocation constraint applies to each state, as opposed to each vaccination site. In other words, the model assumes vaccines can be reallocated between vaccination sites within a state, thus maintaining a degree of freedom in intra-state vaccine distribution.
    \item[--] \textbf{Eligibility}: We prevent people from being vaccinated twice: a patient who has been vaccinated but remains susceptible cannot be vaccinated again. Equation~\eqref{eq:eligibility_constraint} defines the number eligible people as the previous number of eligible people minus the number of people for whom the vaccine was effective and the number of people who got exposed to the disease. Equation~\eqref{eq:eligibility_constraint} then ensures that the number of vaccinated people lies below the number of eligible people.
    \item[--] \textbf{Smoothness}: Large fluctuations in the number of vaccines allocated to each region from day to day would likely cause problems from a supply chain management perspective---both to deliver and to administer the vaccines. Equation~\eqref{eq:smoothness_constraint} ensures that such fluctuations remain minimal. The hyperparameter $\theta_S$ controls the trade-off between efficiency and smoothness.
    \item[--] \textbf{Fairness}: To be politically and socially viable, vaccine distribution must not neglect any region, even if it is not a virus ``hot spot''. This also enhances the robustness of the solution, given that inter-regional transmission can occur in practice. Equation~\eqref{eq:fairness_sites} promotes inter-state fairness at the strategic level, by ensuring that the fraction of vaccination sites in each state does not deviate too much from its population share. Equation~\eqref{eq:fairness_dist} promotes inter-site fairness, by ensuring that vaccine distribution across sites does not deviate too much from uniform distribution. Finally, Equation~\eqref{eq:fairness_vaccines} promotes inter-state fairness at the tactical level, by ensuring that no state receives a fraction of vaccines that exceeds its population share by a wide margin. The hyperparameters $\Theta_L$, $\theta_V$ and $\theta_P$ control the trade-off between efficiency and fairness. As the results will show, even tight fairness constraints leave critical flexibility when locating vaccination sites and allocating vaccines.
    \item[--] \textbf{DELPHI--V dynamics}: Equations~\eqref{eq:susceptible_constraint}--\eqref{eq:immune_constraint} capture the dynamics of the DELPHI--V model in a discretized time space (Equations~\eqref{eq:susceptible_ode}--\eqref{eq:immune_ode}).
    \item[--] \textbf{Domain of definition}: Equation~\eqref{eq:domain_constraint} defines the domain of each variable.
\end{itemize}

\subsubsection*{Model Structure}\ 

Problem $(\mathcal{P})$ is a non-linear program, due to the bilinear terms in Equations~\eqref{eq:susceptible_constraint}--\eqref{eq:exposedV_constraint}, which reflect the fact that the number of new infections result from the interactions between susceptible and infected populations---a key characteristic of all SEIR-based compartmental models. These bilinear terms result in non-convex constraints, thus in a highly challenging optimization model.

The latest Gurobi 9.0 release includes a solver for non-convex quadratic problems \citep{gurobi}. Yet, general-purpose technologies are limited to small-scale instances. In our setting, Problem $(\mathcal{P})$ includes $2m(K+1)T$ non-convex constraints each involving $2(K+1)$ bilinear terms, for a total of $4mT(K+1)^2$ bilinear terms. A realistically-sized problem with $M=51$ (50 US states plus Washington, D.C.), $K=6$ (6 age groups) and $T=90$ (a 3 month planning horizon with daily discretization) would result in nearly 900,000 bilinear terms. Problem $(\mathcal{P})$ remains intractable with existing commercial solvers, motivating the development of a tailored algorithm.

\section{Solution algorithm}
\label{algorithm}

We propose an iterative coordinate descent algorithm to solve Problem $(\mathcal{P})$ in short computational times---consistent with practical requirements. We describe the algorithm in this section. We also present three baselines replicating reasonable strategies that could be implemented in the absence of our data-driven optimization model. These baselines are used for two purposes: (i) to provide an initial feasible solution in the coordinate descent algorithm, and (ii) as benchmarks to evaluate the benefits of the data-driven optimization approach proposed in this paper.

\subsection{Algorithm design}
\label{algorithm_design}

Our algorithm relies on two key observations: 1) aside from Equations~\eqref{eq:susceptible_constraint}--\eqref{eq:exposedV_constraint}, the objective function and all other constraints in $(\mathcal{P})$ are linear, and 2) given a fixed vaccine distribution strategy, the discretized DELPHI--V model can be solved efficiently. Therefore, we proceed by coordinate descent, alternating between two modules: one that optimizes the vaccination distribution strategy given the infection dynamics, and one that simulates the bilinear dynamics of the pandemic for a given vaccination distribution strategy. The optimization part reduces to a linear program, which can be solved very efficiently. Using the resultant vaccine distribution, the simulation part re-estimates the infected population under bilinear dynamics, using a forward discretization scheme.  Specifically, the two modules are defined as follows:
\begin{enumerate}
    \item \textbf{Simulate}: Based on a vaccine allocation solution $\mathbf{V}$, we compute the DELPHI--V dynamics from $t=0$ to $t=T$ (Section~\ref{delphi_with_vaccinations}) by solving the ODE system (Equations~\eqref{eq:susceptible_ode}--\eqref{eq:immune_ode}) using a forward difference scheme in a discretized time space. This terminates in $\mathcal{O}(MKT)$ operations. We denote the total infected population (across all risk classes and vaccinated people) in region $j$ at time $t\in\mathcal{T}$ by $\widehat{I}_{jt}=\sum_{k=1}^K I_{jkt}+I'_{jt}$. We refer to this procedure as $\texttt{Simulate}\left(\mathbf{x},\mathbf{C},\mathbf{W},\mathbf{V}\right)$.
    \item \textbf{Optimize}: Given the infectious population estimates $\mathbf{\widehat{I}}$, we can approximate Equation~\eqref{eq:susceptible_constraint}--\eqref{eq:exposedV_constraint} by the following linear constraints. The problem can then be efficiently solved as a linear programming model. We refer to this module as $\texttt{Optimize}\left(\widehat{\mathbf{I}}\right)$.
    \begin{align*}
    &\quad S_{j,k,t+1} \geq S_{jkt} - \beta V_{jkt} - \alpha_j\gamma_{jt}\left(S_{jkt}-\beta V_{jkt}\right)\widehat{I}_{jt}\Delta t, && \forall j \in \mathcal{M}, k \in \mathcal{K}, t \in \mathcal{T} \\
    &\quad S'_{j,t+1} \geq S'_{jt} + \beta \sum_{\kappa=1}^KV_{j\kappa t} - \alpha_j\gamma_{jt}\left(S'_{jt} + \beta \sum_{\kappa=1}^KV_{j\kappa t}\right)\widehat{I}_{jt}\Delta t, && \forall j \in \mathcal{M}, t \in \mathcal{T} \\
    &\quad E_{j,k,t+1} \geq E_{jkt} +\left( \alpha_j\gamma_{jt}\left(S_{jkt}-\beta V_{jkt}\right)\widehat{I}_{jt} - r^I E_{jkt}\right)\Delta t, && \forall j \in \mathcal{M},\  k \in \mathcal{K},\  t \in \mathcal{T}  \\
    \label{eq:}
    &\quad E'_{j,t+1} \geq E'_{jt} + \left( \alpha_j\gamma_{jt}\left(S'_{jt} + \beta \sum_{\kappa=1}^KV_{j\kappa t}\right)\widehat{I}_{jt} - r^I E'_{jt}\right)\Delta t, && \forall j \in \mathcal{M}, t \in \mathcal{T}
    \end{align*}
\end{enumerate}

We iterate between the $\texttt{Simulate}$ and $\texttt{Optimize}$ modules, until convergence. Specifically, the algorithm terminates when the variation in the objective function value remains minimal from one iteration to the next. The pseudocode summarizing this approach is presented in Algorithm~\ref{alg:coordinate_descent}. We turn next to the generation of an initial feasible solution.

\begin{algorithm}[h!]
\SetKwInOut{Input}{input}
\SetKwInOut{Output}{output}
\Input{Prescriptive DELPHI--V--OPT data; termination tolerance $\varepsilon$ \\
}
    Initialization: $i \gets 0$,\  $\mathbf{x}^{(i)}\gets\mathbf{0},\  ,\mathbf{C}^{(i)}\gets\mathbf{0},\  ,\mathbf{W}^{(i)}\gets\mathbf{0},\  ,\mathbf{V}^{i}\gets\mathbf{0},\  \mathbf{I}^{i}\gets\mathbf{0},\ Z^{i}\gets\infty$ \\
    $i\gets i + 1$ \\
    $\left(\mathbf{x}^{(i)},\mathbf{C}^{(i)},\mathbf{W}^{(i)},\mathbf{V}^{(i)}, \mathbf{I}^{(i)}, Z^{(i)}\right) \gets \texttt{GenerateFeasibleSolution}$ \\
    \While{$\left|Z^{(i)} - Z^{(i-1)}\right|/Z^{(i-1)} > \varepsilon'$} {
    $i\gets i + 1$ \\
    Run $\texttt{Simulate}\left(\mathbf{x}^{(i-1)},\mathbf{C}^{(i-1)},\mathbf{W}^{(i-1)},\mathbf{V}^{(i-1)}\right)$. Update $\widehat{\mathbf{I}}^{(i)} \gets \widehat{\mathbf{I}}$, where $\widehat{\mathbf{I}}$ is the output of $\texttt{Simulate}\left(\mathbf{x}^{(i-1)},\mathbf{C}^{(i-1)},\mathbf{W}^{(i-1)},\mathbf{V}^{(i-1)}\right)$. \\
    Run $\texttt{Optimize}\left(\widehat{\mathbf{I}}^{(i)}\right)$. Update $\left(\mathbf{x}^{(i)},\mathbf{C}^{(i)},\mathbf{W}^{(i)},\mathbf{V}^{(i)}\right)\gets \left(\mathbf{x},\mathbf{C},\mathbf{W},\mathbf{V}\right)$, where $\left(\mathbf{x},\mathbf{C},\mathbf{W},\mathbf{V}\right)$ is the output of $\texttt{Optimize}\left(\widehat{\mathbf{I}}^{(i)}\right)$. Update the objective function:
    $$Z^{(i)}\gets\sum_{j=1}^{M}\sum_{k=1}^K \left(D_{jkT} + H_{jkT} + Q_{jkT}\right)+\lambda_E\sum_{j=1}^{M}\sum_{k=1}^KE_{jkT}+\lambda_D\sum_{l=1}^L\sum_{i=1}^nPop_l\Delta_{li}W_{li}$$
    }
\Output{Vaccine distribution strategy $\left(\mathbf{x}^{(i)},\mathbf{C}^{(i)},\mathbf{W}^{(i)},\mathbf{V}^{(i)}\right)$}
\caption{Coordinate descent algorithm for DELPHI--V--OPT (Problem $(\mathcal{P})$).}
\label{alg:coordinate_descent}
\end{algorithm}

\subsection{Baselines}
\label{subsec:baselines}

We propose three simple and interpretable baselines for generating a feasible solution to $(\mathcal{P})$. By design, these baselines are heuristics that solely rely on the inputs of the optimization models, as opposed to requiring the full model and algorithm developed in this paper.

\noindent
\textbf{Top-cities baseline}: This approach prioritizes cities based on population. Specifically, it deploys vaccination sites in the most populous cities, while accounting for the constraint that each state must have at least one center (Equation~\eqref{eq:one_per_state}). Subsequently, it allocates an equal fraction of the daily vaccine budget to each vaccination site: we fix the variables $\mathbf{x}$ and $\mathbf{C}$, run the model to optimize vaccine allocation within each state, and estimate the resulting number of deaths. This baseline corresponds to a city-level approach based on demographic information alone.

\noindent
\textbf{Population-based baseline}: Under this approach, the number of vaccination sites deployed in each state is based on the state's population share. This is formulated as follows, where $y_j$ is a decision variable denoting the number of vaccination sites in state $j$.
$$\min\quad\sum_{j=1}^m\left|\frac{\sum_{l\in\mathcal{L}_j}Pop_l}{\sum_{j'=1}^m\sum_{l\in\mathcal{L}_{j'}}Pop_l}N-y_j\right|,\ \text{s.t. $\sum_{j=1}^my_j=100$,\ $y_j\geq1,\forall j\in\mathcal{M}$,\ $\mathbf{y}$ integer.}$$
We then solve DELPHI--V--OPT, while fixing the aggregate number of vaccination sites per state, i.e., $\sum_{i\in\mathcal{N}_j}x_i=y_j$, and assuming equal allocation of vaccines across vaccination sites, i.e., $C_{it}=\frac{1}{N}B_t,\ \forall t\in\mathcal{T}$. The model allocates vaccines within each state and estimates the number of deaths. This baseline corresponds to a state-level approach based on demographic information alone.

\noindent
\textbf{Case-based baseline}: Under this approach, the number of vaccination sites deployed in each state is based on the number of COVID-19 cases at the beginning of the planning horizon. This is formulated as follows, where $y_j$ is a decision variable denoting the number of vaccination sites in state $j$ and $Cases_j$ denotes the case count in state $j$.
$$\min\quad\sum_{j=1}^m\left|\frac{Cases_j}{\sum_{j'=1}^mCases_{j'}}N-y_j\right|,\ \text{s.t. $\sum_{j=1}^my_j=100$,\ $y_j\geq1,\forall j\in\mathcal{M}$,\ $\mathbf{y}$ integer.}$$
We then proceed as with the population-based baseline, by fixing the number of vaccination sites per state, assuming equal vaccine allocation across sites, and re-solving the model. This baseline corresponds to a state-level approach based on epidemiological information alone.

By design, these baselines satisfy all constraints of Problem $(\mathcal{P})$, and thus provide valid initializations into our coordinate descent algorithm. They also provide sensible and equitable benchmarks based on readily-available demographic information (e.g., census data) and epidemiological information (e.g., case counts), hence easily implementable. Comparisons between our optimized solution and these benchmarks thus estimate the benefits of vaccine distribution optimization.

\section{Experimental setup}
\label{methodology}

We implement the proposed model and algorithm in the United States. We select $N=100$ vaccination sites out of the 500 most populous cities in the United States as candidate locations (set $\mathcal{N}$) We define the set $\mathcal{M}$ as 51 ``states'' (the 50 states plus the District of Columbia) and the set $\mathcal{L}$ as the 3,006 counties. We define six risk classes based on six relatively coarse age groups: 0-9 years, 10-49 years, 50-59 years, 60-69 years, 70-79 years, and 80 years and above. These simplified risk classes facilitate the practical implementation of the solution while capturing broad trends in mortality rates. We define the time horizon $\mathcal{T}$ as the three-month period from February to April 2021, consistently with the ongoing planning horizon of the US federal government.

\subsection{Data sources}

\begin{figure*}[h!]
\centering
\begin{tikzpicture}[scale=0.55]
\node[rounded rectangle, draw=myblue, text width=3cm, align=center,draw] (NYT) at (0, 11) {NYT data};
\node[rounded rectangle, draw=myblue, text width=3cm, align=center] (Census) at (10, 11) {US census data};
\node[rounded rectangle, draw=myblue, text width=3cm, align=center] (CDC) at (20, 11) {CDC data};
\node[rounded rectangle, draw=myred, text width=4cm, align=center] (DELPHI params) at (5, 7) {DELPHI parameters};
\node[rounded rectangle, draw=myred, text width=4cm, align=center] (Population) at (15, 7) {Population by region and risk class};
\node[rounded rectangle, draw=black, text width=4cm, align=center] (Mortality optimization) at (20, 2) {Mortality rate estimation};
\node[rounded rectangle, draw=black, text width=4cm, align=center] (DELPHI) at (7, 2) {Trained DELPHI\\(without vaccinations)};
\node[rounded rectangle, draw=myred, text width=4cm, align=center] (DELPHI preds) at (7, -3) {Initial conditions (from DELPHI)};
\node[rounded rectangle, draw=myred, text width=4cm, align=center] (Mortality rates) at (20, -3) {Mortality rates by region and risk class};
\node[rounded rectangle, draw=black, text width=4cm, align=center] (DELPHI model) at (10, -8) {Prescriptive model: DELPHI--V--OPT};
\node[rounded rectangle, draw=mygreen, text width=4cm, align=center] (Policy) at (10, -12) {Vaccine distribution strategy};
\draw[->] (NYT)--(DELPHI params);
\draw[->] (Census)--(Population);
\draw[->] (DELPHI params)--(DELPHI);
\draw[->] (DELPHI)--(DELPHI preds);
\draw[->] (Population) to [bend left] (DELPHI preds);
\draw[->] (DELPHI)--(Mortality optimization);
\draw[->] (CDC)--(Mortality optimization);
\draw[->] (Mortality optimization)--(Mortality rates);
\draw[->] (Population)--(Mortality optimization);
\draw[->] (DELPHI preds)--(DELPHI model);
\draw[->] (DELPHI params) to [bend right=75] (DELPHI model);
\draw[->] (Mortality rates)--(DELPHI model);
\draw[->] (DELPHI model)--(Policy);
% \node[draw, dotted, fit=(JHU) (Census) (CDC)] {}
% \node[draw, dotted, fit=(DELPHI params) (DELPHI preds) (Population) (Mortality rates)] {}
\end{tikzpicture}
\caption{Simulation environment: raw data (blue), processed data (red), models (black) and outputs (green).}
\label{fig:data_flowchart}
\end{figure*}
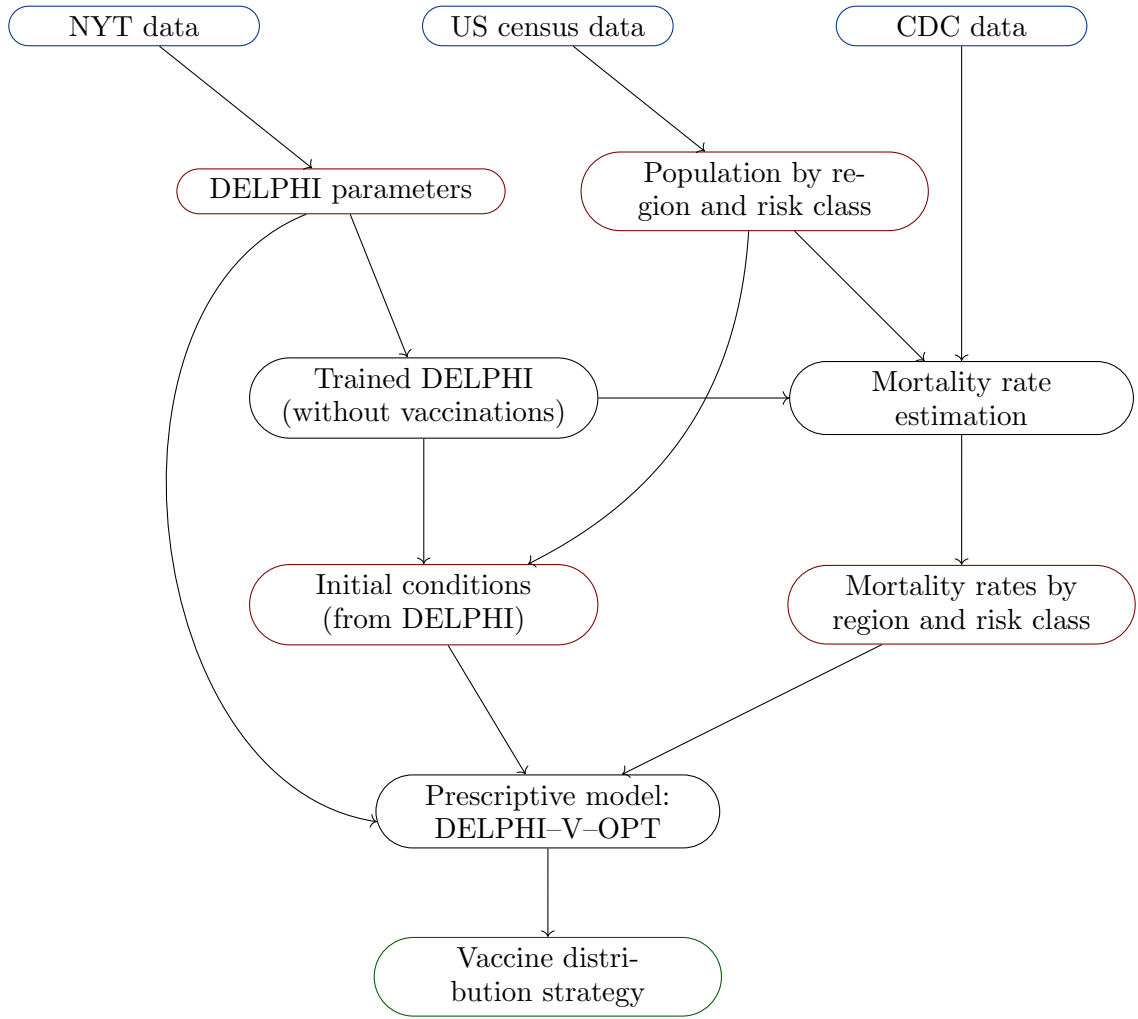

We calibrate the model using multiple data sources (Figure~\ref{fig:data_flowchart}). First, we estimate the parameters of the DELPHI model (without vaccinations) independently for each state, using historical data on cases and deaths from the \cite{nyt:20}. We obtain a granular population breakdown by age for each state from the \cite{uscensus:01}. We then run DELPHI (still, without vaccinations) to derive the initial susceptible, exposed and infected populations (on January 30, 2021), which we distribute among the risk classes proportionally to their size.

The next input is an estimate of the mortality rate per region, risk class and time period. We make use of the data from the \cite{cdc:death_count}, which report the total number of confirmed COVID-19 cases, hospitalizations and deaths by age group in the United States until the end of January 2021. In contrast, the DELPHI model fits a time-dependent mortality rate within each region. To the best of our knowledge, there exists no other data source for nationwide cases and deaths by age group. This leaves a discrepancy between the time-independent estimates at the risk class level from the CDC, and the time-varying estimates at the region-level from DELPHI. To reconcile these data, we employ an optimization procedure that interpolates the mortality rate per region, risk class and time period. We present this approach in the next section.

\subsection{Mortality rate estimation}
\label{mortality_rate_calibration}

Our procedure to estimate mortality rates starts from two sets of inputs:
\begin{itemize}
    \item \textbf{DELPHI predictions:} Let $\widehat{X}_{jt}$ denote the estimated number of new detected cases in region $j\in\mathcal{M}$ and time period $t\in\mathcal{T}$. Let $\widehat{D}^d_{j,t+\tau_j}$ denote the number of deaths, where $\tau_j$ is the median death time after detection in region $j$. These quantities are aggregated across risk classes.
    \item \textbf{CDC data:} Let $X^{\text{CDC}}_k$ and $D^{\text{CDC}}_k$ denote the number of cases and deaths for risk class $k\in\mathcal{K}$. These quantities are aggregated across regions and time periods.
\end{itemize}

We define the reference mortality rate $\overline{m}_{jk}$ of risk class $k\in\mathcal{K}$ based in region $j$ as follows:
\begin{equation}\label{eq:reference}
    \overline{m}_{jk} = \frac{D^{\text{CDC}}_k}{X^{\text{CDC}}_k}\left(\frac{\sum_{t=1}^T\widehat{D}^d_{j,t+\tau_j}}{\sum_{t=1}^T\widehat{X}_t}\right)\left(\frac{\sum_{l=1}^K X^{\text{CDC}}_l}{\sum_{l=1}^KD^{\text{CDC}}_l}\right)
\end{equation}
By design, this expression preserves the ratio of mortality rates between different risk classes from the CDC data, while correcting the mean reference mortality rate in each region across the planning horizon to be in line with the DELPHI projections.

We then estimate the mortality rate for each region $j\in\mathcal{M}$, risk class $k\in\mathcal{K}$, and time period $t\in\mathcal{T}$, denoted by $m_{jkt}$. We also introduce additional decision variables $X_{jkt}$ and $D_{j,k,t+\tau_j}$, reflecting the number of detected cases and the number of future deaths in region $j$ assigned to risk class $k\in\mathcal{K}$ and time period $t\in\mathcal{T}$. Given that the fitting procedure is done separately in each region $j\in\mathcal{M}$, we decouple the problem at the region level---thereby considerably reducing the size of each problem instance. Specifically, we formulate the optimization problem given in Equations~\eqref{eq:obj_mortality}--\eqref{eq:constr_8_mort}.
\begin{align}
    \label{eq:obj_mortality}
    \text{for all $j \in \mathcal{M}$:}\ \min ~& \sum_{t=1}^T\sum_{k=1}^K \left[\left(\frac{m_{jkt} - \overline{m}_{jk}}{\overline{m}_{jk}}\right)^2 + \lambda_M\left|\frac{X_{jkt} - p_{jk}\widehat{X}_{jt}}{p_{jk}\widehat{X}_{jt}}\right|\right] \\
    \label{eq:constr_2_mort}
    \text{s.t.} ~& \sum_{k=1}^K D_{j,k,t+\tau_j} = \widehat{D}_{j,t+\tau_j}, && \forall t \in \mathcal{T} \\
    \label{eq:constr_3_mort}
    & \sum_{k=1}^K X_{jkt} = \widehat{X}_{jt}, && \forall t \in \mathcal{T} \\
    \label{eq:constr_1_mort}
    & m_{jkt} \cdot X_{jkt} = D_{j,k,t+\tau_j}, &&  \forall k \in \mathcal{K}, \forall t \in \mathcal{T},\\
    \label{eq:constr_5_mort}
    & m_{jkt} \geq m_{j,k,t+1}, &&  \forall k \in \mathcal{K}, \forall t \in \mathcal{T} \\ 
    \label{eq:constr_6_mort}
    & m_{jkt}  \geq m_{jlt},  &&  \forall k,l \in \mathcal{K}:  \overline{m}_k \geq \overline{m}_l,\\
    \label{eq:constr_7_mort}
    &  m_{jkt} \leq 1, &&  \forall k \in \mathcal{K}, \forall t \in \mathcal{T} \\
    \label{eq:constr_8_mort}
    &\mathbf{m}, \mathbf{C}, \mathbf{D} \geq \mathbf{0}.
\end{align}

The first term in Equation~\eqref{eq:obj_mortality} minimizes the squared relative error between the mortality rate estimates and their reference values (Equation~\eqref{eq:reference}). The second term is a regularization penalty that minimizes deviations between the proportion of detected cases and the proportion $p_{jk}$ of the population, in each risk class. The parameter $\lambda_M$ trades off these two objectives (we use $\lambda_M=0.1$ in our experiments). Equations~\eqref{eq:constr_2_mort}--\eqref{eq:constr_3_mort} ensure consistency with the DELPHI predictions. Equation~\eqref{eq:constr_1_mort} defines the mortality rate as the ratio between the number of deaths and cases. Equations~\eqref{eq:constr_5_mort}--\eqref{eq:constr_6_mort} ensure that mortality rates are decreasing over time and monotonic with risk classes. Finally, Equations~\eqref{eq:constr_7_mort}--\eqref{eq:constr_8_mort} define the domain of the variables.

Note that the problem is non-linear due to the bilinear term in Equation~\eqref{eq:constr_1_mort}. Yet, thanks to the decoupling at the region level, we can solve the problem using the quadratic solver in Gurobi 9.0 \citep{gurobi}, which addresses non-convexities using branching and cutting planes algorithms. In practice, a solution within a 1\% optimality gap is generally obtained within minutes.

The output of this algorithm is an estimate of the mortality rate at the level of each state and each risk class, in each time period. We report aggregated statistics in Table~\ref{tab:mortality_rates}.

\begin{table*}[h!]
\centering
\begin{tabular}{@{}lrrrrrr@{}}
\toprule
Month & 0-9 & 10-49 & 50-59 & 60-69 & 70-79 & 80+ \\ \midrule
February & 0.008\% & 0.119\% & 0.882\% & 2.271\% & 6.101\% & 15.027\% \\
March & 0.007\% & 0.116\% & 0.859\% & 2.212\% & 5.942\% & 14.636\% \\
April & 0.007\% & 0.114\% & 0.844\% & 2.171\% & 5.832\% & 14.365\% \\ \bottomrule
\end{tabular}
\caption{Calibrated monthly mortality rates, averaged by risk class and over all states.}
\label{tab:mortality_rates}
\end{table*}

\subsection{Implementation details}
\label{implementation}

Our DELPHI--V--OPT model relies on a forward difference scheme to simulate the dynamics of the pandemic, for any vaccine allocation. If the discretization is too coarse, the algorithm will introduce truncation errors. If, however, the discretization is too granular, computational times will be prohibitively long. After extensive experimentations, we found $\Delta t = 1$ day to yield the best compromise. It is also practical choice as it yields a day-by-day plan.

The coordinate descent scheme terminates when the change in the objective function value lies below a pre-determined threshold. We set the termination tolerance to $\varepsilon=0.1\%$.

Regarding DELPHI-V--OPT, we vary the hyperparameters $\lambda_E$, $\lambda_V$, $\theta_S$, $\Theta_L$, $\theta_V$ and $\theta_P$ to balance efficiency with practical and fairness considerations, and perform sensitivity analyses on these parameters. We consider a baseline vaccine effectiveness of 90\% (in line with the values from the first vaccine approvals) and a baseline budget of 1 million vaccines per day (which can be viewed as 1 million single-dose vaccines, 2 million double-dose vaccines or a combination thereof). Given the strong underlying uncertainty, we perform sensitivity analyses to demonstrate the robustness of the benefits of our optimization outputs to vaccine effectiveness and vaccine budget.

All optimization models are implemented in Gurobi 9.0, with 2.3GHz processor and 4 cores. We use a barrier method to solve each linear program, with a barrier convergence tolerance of $10^{-6}$. 

\section{Experimental results}
\label{results}

We now present results obtained with the modeling and algorithmic framework developed in this paper. Our main focus is on the location of the 100 vaccination sites---decisions that have to be made immediately---as opposed to vaccine allocation---decisions that can be revisited continuously over time as more information becomes available. We first evaluate the benefits of the optimized solution against baseline approaches, and then establish the robustness of these benefits.

\subsection{Benefits of optimizing the vaccine distribution strategy}

\begin{figure*}[h!]
    \centering
    \includegraphics[width=.8\textwidth]{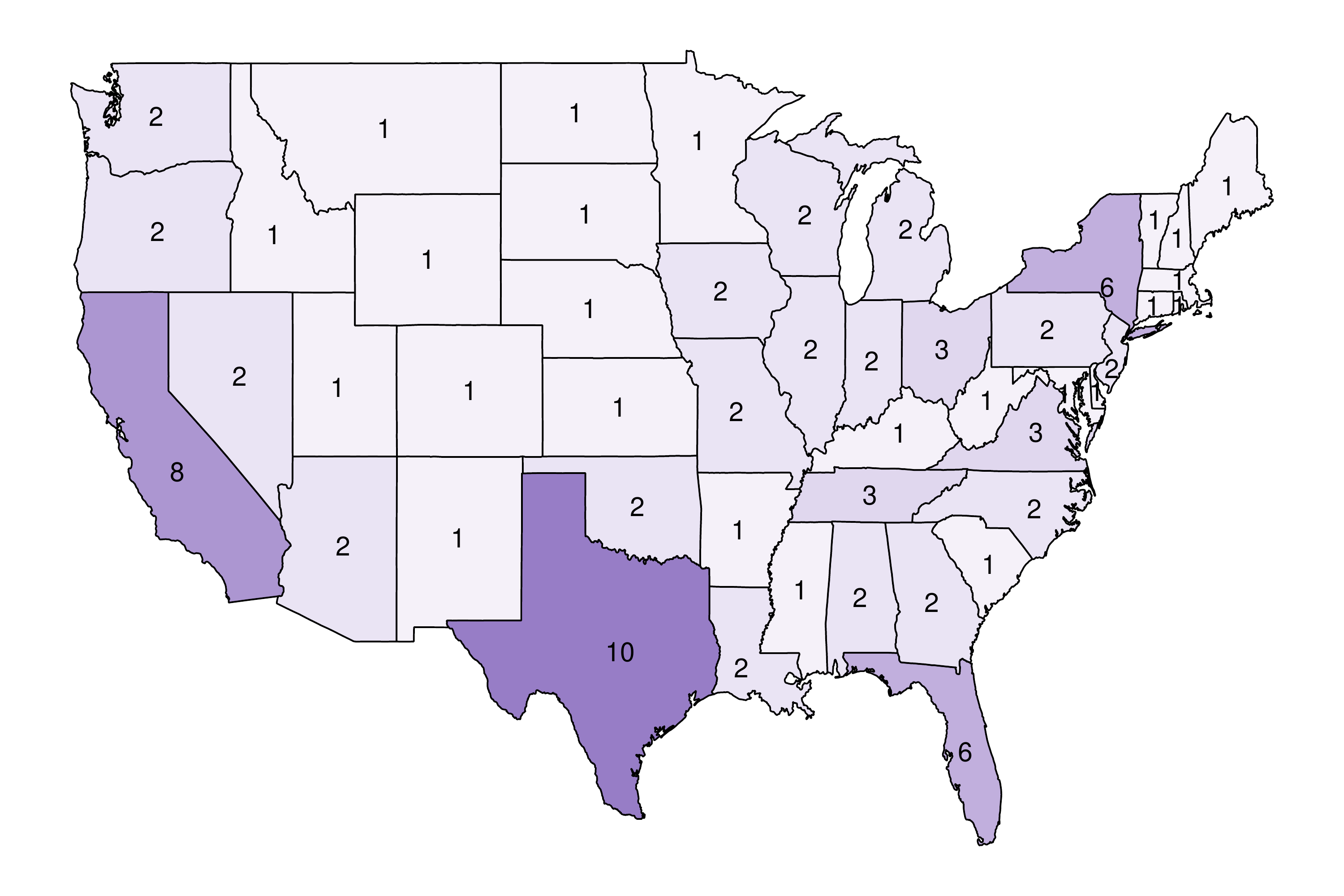}
    \caption{Number of centers per state in the proposed solution.}
    \label{fig:map_count}
\end{figure*}

Figure~\ref{fig:map_count} shows the heatmap of vaccination sites across all 51 states (the exact list of proposed locations is reported in the appendix). Recall that, by design, the optimized solution deploys at least one vaccination site per state (Equation~\eqref{eq:one_per_state}). In addition, one of the fairness constraints (Equation~\eqref{eq:fairness_sites} anchors the number of vaccination sites per state to its population share. Other than that, the vaccination sites are deployed strategically to curb the spread of the pandemic. As a result, the optimized number of vaccination sites varies significantly per state, as a function of underlying demographic and epidemiological factors. For instance, the four largest states by population (California, Texas, Florida and New York) receive the most vaccination sites (6 to 10 each). In contrast, the smaller states receive only one vaccination site. Ultimately, the heatmap suggests that the optimized solution targets large population centers and some of the hot spots of the pandemic, while ensuring equity nationwide.

Then, could we have achieved a similar solution with some of our benchmarks (Section~\ref{subsec:baselines}), which determine the locations of the vaccination sites based on demographic and epidemiological data but do not make use of our epidemiological and optimization models? To investigate this question, we evaluate the dynamics of the pandemic under each of the three baselines. When it comes to the optimization, we derive three solutions: (i) an ``optimized locations'' solution, which optimizes the site locations decisions freely (without fairness constraints) but then allocates vaccines uniformly across the 100 selected vaccination sites; (ii) a ``fully optimized'' solution, which optimizes the site locations decisions and vaccine allocation decisions freely (without fairness constraints); and (iii) a ``proposed'' solution, which optimizes the site locations decisions and vaccine allocation decisions with fairness constraints. For each solution, Table~\ref{tab:saved} reports the number of lives saved by the vaccination campaign, as compared to a solution without vaccinations.

\begin{table}[h]
    \centering
    \renewcommand{\arraystretch}{1.2}
    \caption{Comparison of optimized solutions and benchmarks.}
    \label{tab:saved}
    \footnotesize{
        \begin{tabular}{ccccc}
        \toprule\toprule
            Method & Solution & Site locations & Vaccine distribution & Saved lives \\
        \midrule
            &&  &     &   19,045   \\
            &\multirow{-2}{*}{Top-cities}& \multirow{-2}{*}{Most populous cities}&  \multirow{-2}{*}{Uniform across centers} &    (base)   \\
            \cmidrule{2-5}
            &&  &     &   19,709   \\
            &\multirow{-2}{*}{Population}& \multirow{-2}{*}{Pro-rata population}&  \multirow{-2}{*}{Uniform across centers} &    (+3.5\%)   \\
            \cmidrule{2-5}
            &&  &     &   21,037   \\
            \multirow{-7}{*}{Benchmarks} & \multirow{-2}{*}{Cases}& \multirow{-2}{*}{Pro-rata active cases}&  \multirow{-2}{*}{Uniform across centers} &  (+10.5\%)   \\
            \midrule
            &&  &     &   23,622   \\
            &\multirow{-2}{*}{Locations}& \multirow{-2}{*}{Optimized: minimizes deaths}&  \multirow{-2}{*}{Uniform across centers} &   (+24.0\%)   \\
            \cmidrule{2-5}
            &&  &     &   25,615   \\
            &\multirow{-2}{*}{Full}& \multirow{-2}{*}{Optimized: minimizes deaths}&  \multirow{-2}{*}{Optimized: minimizes deaths} &   (+34.5\%)   \\
            \cmidrule{2-5}
            && Optimized: minimizes deaths &  Optimized: minimizes deaths   &   23,000   \\
            \multirow{-7}{*}{Optimization}&\multirow{-2}{*}{Proposed}& Fair: inter-state equity &  Fair: inter-center equity &   (+20.8\%)   \\
        \bottomrule\bottomrule
        \end{tabular}
    }
\end{table}

Let us start with the main observation: the optimized solutions provides significant benefits, as compared to all benchmarks. Comparing the ``optimized locations'' solution to the top-cities benchmark, we find that optimizing locations alone can save around 4,500 lives over the three-month period under consideration, enhancing the impact of the vaccination campaign by 24\%. Moving to the ``fully optimized'' solution, we note that jointly optimizing locations and vaccine allocation achieves further benefits, by saving an extra 2,000 lives and increasing the benefits over the benchmark to 35\%. These results underscore the potential of the proposed optimization approach, which leverages available vaccines strategically to target the regions that need them most. Another observation is that determining the locations of vaccination sites does not achieve all potential benefits of the vaccination campaign, underscoring the role of downstream vaccine allocation as an extra lever to combat the pandemic.

A downside of these two solutions (``locations optimized'' and ``fully optimized''), however, is that they can result in very inequitable outcomes between states and between vaccination sites. The last (``proposed'') solution circumvents this challenge by imposing all fairness constraints (Equations~\eqref{eq:fairness_sites}--\eqref{eq:fairness_vaccines}), and choosing tight values for the corresponding hyperparameters $\Theta_L$, $\theta_V$ and $\theta_P$. As such, the proposed solution ensures fairness across various dimensions (site locations, vaccine allocation across states, and vaccine allocation across sites). Remarkably, even when constraining the optimization as such, the resulting proposed solution still results in 20\% benefits, as compared to the top-cities benchmark---saving 4,000 deaths over the three-month horizon.

In comparison, the benchmarks cannot reach the same impact of vaccinations. The top-cities and population-based benchmarks achieve similar outcomes, with 19,000--20,000 lives saved as compared to a no-vaccination baseline. The case-based baseline performs better, by saving 21,000 lives (10\% more than the top-cities baseline). Yet, these numbers remain significantly lower than those achieved with our solution. The top-cities and population-based baselines, by relying on demographic information exclusively, fail account for disparities in the severity of the across the country. The case-based baseline captures the status of the pandemic, but not its dynamics, thus treating similarly a state where the pandemic is already waning and a state where it is rising fast---although vaccinations have a stronger impact in the latter state than the former.

These results underscore the main takeaways from this paper: optimization provides significant benefits, as compared to simple benchmarks based on readily-available information at the outset of the vaccination campaign. Instead, the optimized approaches design vaccine distribution strategies based on both demographics and epidemiological dynamics. The edge of optimization can be significant, by saving an extra 20\% of lives with the same vaccine capacity. Stated differently, under the proposed optimization approach, each vaccine is effectively ``worth'' 1.2 vaccines.

\begin{figure}[h!]
    \centering
	\subfloat[Vaccines allocation]{\label{subfig:map_allocation}\includegraphics[width=.5\textwidth]{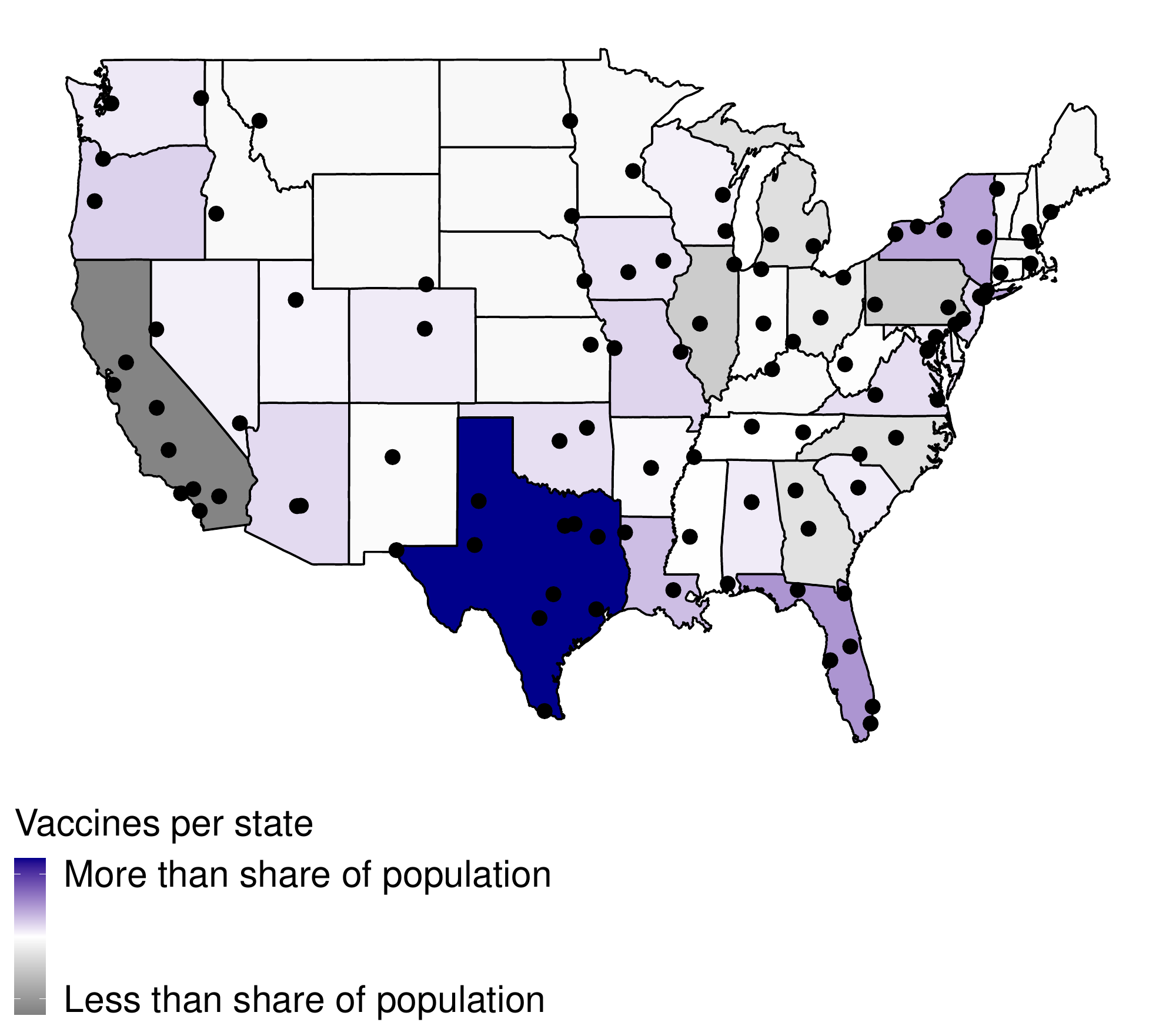}}
	\subfloat[Lives saved]{\label{subfig:map_deaths}\includegraphics[width=.5\textwidth]{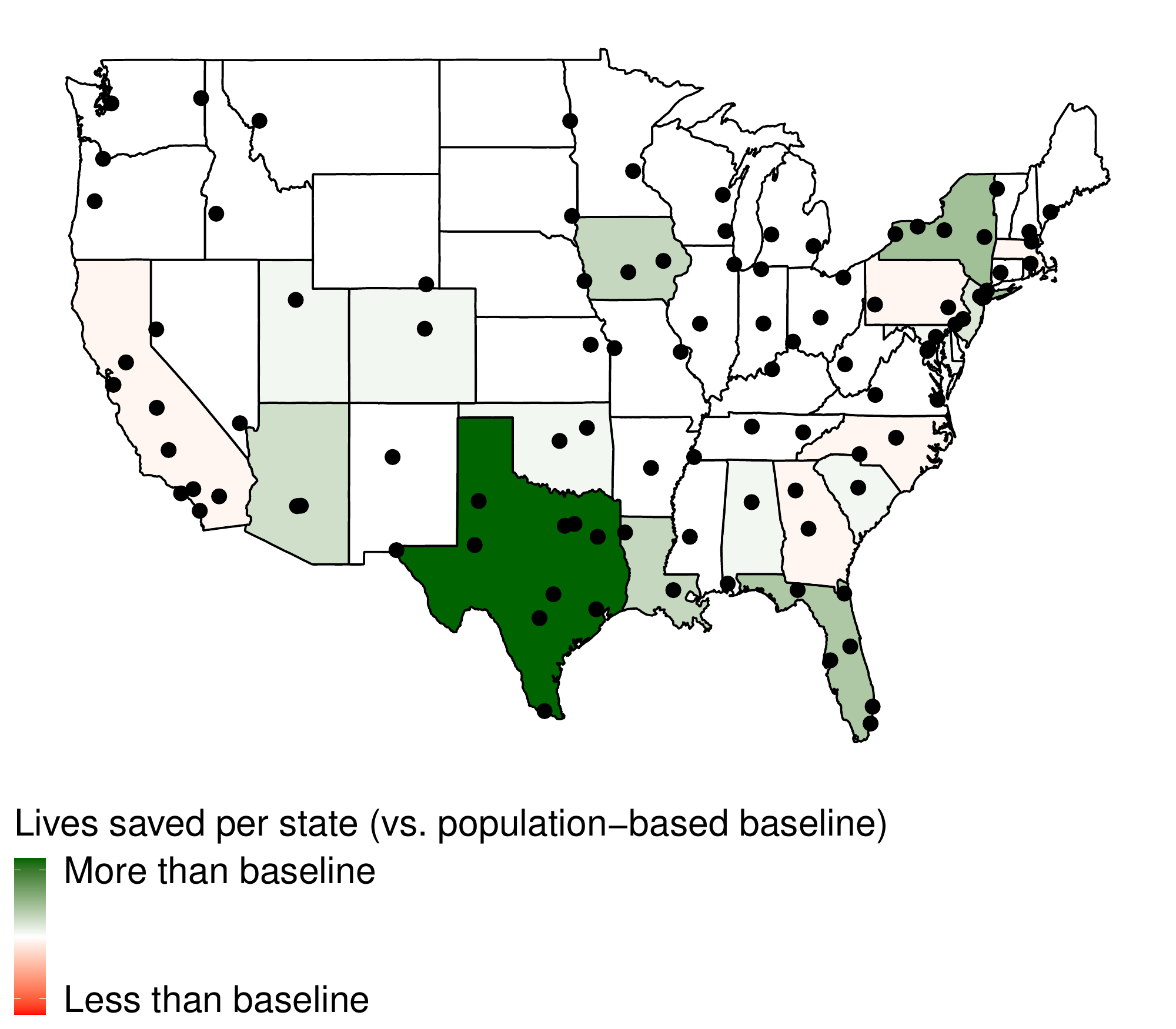}}
	\vspace{6 pt}
	\caption{Vaccine allocation (vs. population share) and lives saved (vs. population-based baseline) per state.}
	\label{fig:map_population_deaths}
	\vspace{-12 pt}
\end{figure}

A natural question to ask is whether optimization induces strong geographic disparities, by merely shifting the benefits of the vaccines from one state to another. To explore this question, Figure~\ref{fig:map_population_deaths} plots the number of vaccines distributed to each state, as compared to its population share (Figure~\ref{subfig:map_allocation}) and the number of lives saved, as compared to the top-cities benchmark (Figure~\ref{subfig:map_deaths}).

Recall that the optimization imposes fairness constraints in vaccine allocation (Equations~\eqref{eq:fairness_sites}--\eqref{eq:fairness_vaccines}); yet, the remaining flexibility can be used strategically to target the populations where vaccines can have the strongest impact. In fact, as Figure~\ref{subfig:map_allocation} shows, vaccines do not get distributed proportionally to each state's population. For instance, states like Texas, Florida and New York receive a higher share of vaccines, whereas states like California, Pennsylvania and Illinois receive a lower share. This is expected, given the significantly higher number of lives saved under the optimized (``proposed'') solution as compared to the population-based benchmark (Table~\ref{tab:saved}).

However, Figure~\ref{subfig:map_deaths} shows that these disparities in vaccine allocation do not result in sharp disparities in public health outcomes. Specifically, the optimized solution saves hundreds of extra lives (as compared to the population-based benchmark) in seven states with very different epidemiological profiles. Texas benefits the most from optimization (with an additional 1,450 lives saved), followed by New York (440), Florida (380) and Iowa (290). At the same time, the optimized solution does not increase the death toll in any state by more than 100. Pennsylvania is the most negatively impacted state, with an estimated 85 additional deaths---well within the margin of error of DELPHI--V. Overall, the proposed optimization approach can thus distribute vaccine capacities to combat the pandemic in some critical states without hurting others.

\begin{edit}
The obvious question then is: how does the proposed solution compare to the vaccination plan that was applied in practice? Unfortunately, it is hard to perform a complete apples-to-apples comparison.  Indeed, the deployment of vaccination site was conducted in a more ad hoc fashion in practice than modeled in this paper, based on local capacity, heterogeneity across “big” and “small” sites, etc. Moreover, since this paper was written, the Federal government has updated its plan, so that the Federal facilities take a supporting role while delegating most authority to state-driven vaccination efforts. Therefore, the full plan was not implemented and instead FEMA has opened just 24 facilities in 13 states, as compared to the original 100 planned ones. We provide details in Table \ref{tab:realized_centers} in the Appendix.

Nonetheless, we provide two sources of evidence supporting the results from our analysis. First, we observe that FEMA allocated a higher share to states such as Florida than would be otherwise allocated on a population pro rata basis. Conversely, other states like California where underweighted. This is consistent with the main recommendations outlined in the paper. Second, even though our recommendations were not applied directly to the location of vaccination centers, the subsequent allocation of vaccines to each state was influenced by the model's recommendations. This underscores a second tactical lever in our model-based recommendation---vaccine allocation---beyond its main strategic lever---location of vaccination sites.
\end{edit}

\subsection{Sensitivity and robustness}

A core challenge in vaccine distribution lies in the significant uncertainty regarding the dynamics of the pandemic and the effect of vaccinations. To address this challenge, we assess the sensitivity and the robustness of the optimized solution when the structure of the DELPHI--V model and some of its key parameters are perturbed. For each perturbation, we compare three solutions:
\begin{itemize}
    \item[--] the top-cities baseline, evaluated with the new perturbed model
    \item[--] the re-optimized solution, optimized and evaluated with the new perturbed model
    \item[--] the proposed solution, optimized with the original model and evaluated with the new perturbed model. Specifically, we first run the optimization with the initial inputs. We then introduce perturbations, and re-optimize vaccine allocation decisions (variables $\mathbf{C}$ and $\mathbf{V}$), while fixing the vaccination sites (variable $\mathbf{x}$). Indeed, in practice, vaccination sites are determined once and for all, whereas vaccine allocation can be re-adjusted as information becomes available.
\end{itemize}
We then compare the number of deaths under these three solutions, estimated with the perturbed DELPHI--V model. Comparisons between the top-cities benchmark and the re-optimized solution estimate the sensitivity of the benefits of optimization to the perturbations. Comparisons between the top-cities benchmark and the proposed solution estimate the robustness of the solution.

\begin{figure}[h!]
    \centering
	\subfloat[Vaccine effectiveness]{\label{subfig:effectiveness}\includegraphics[width=.5\textwidth]{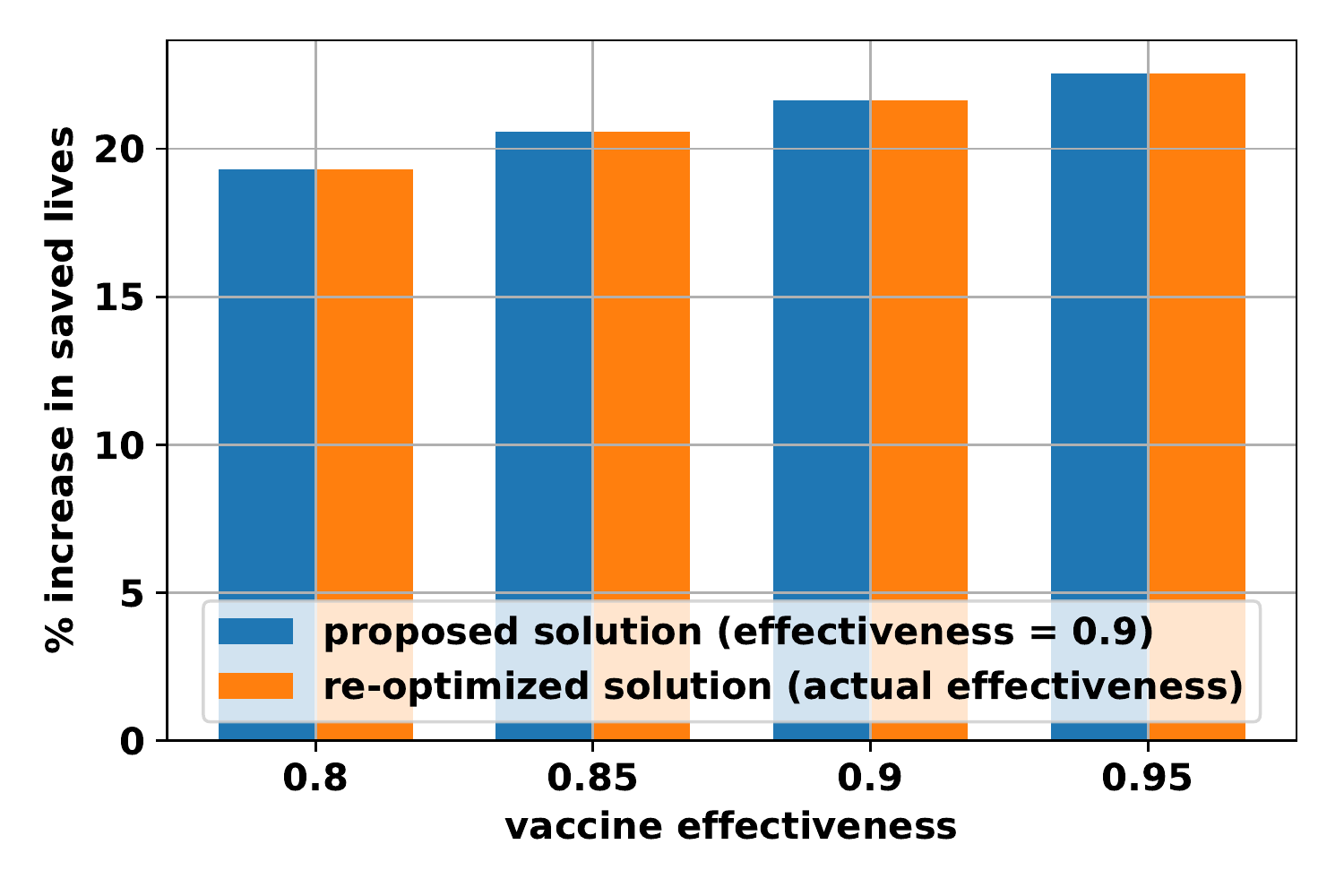}}
	\subfloat[Vaccine budget]{\label{subfig:budget}\includegraphics[width=.5\textwidth]{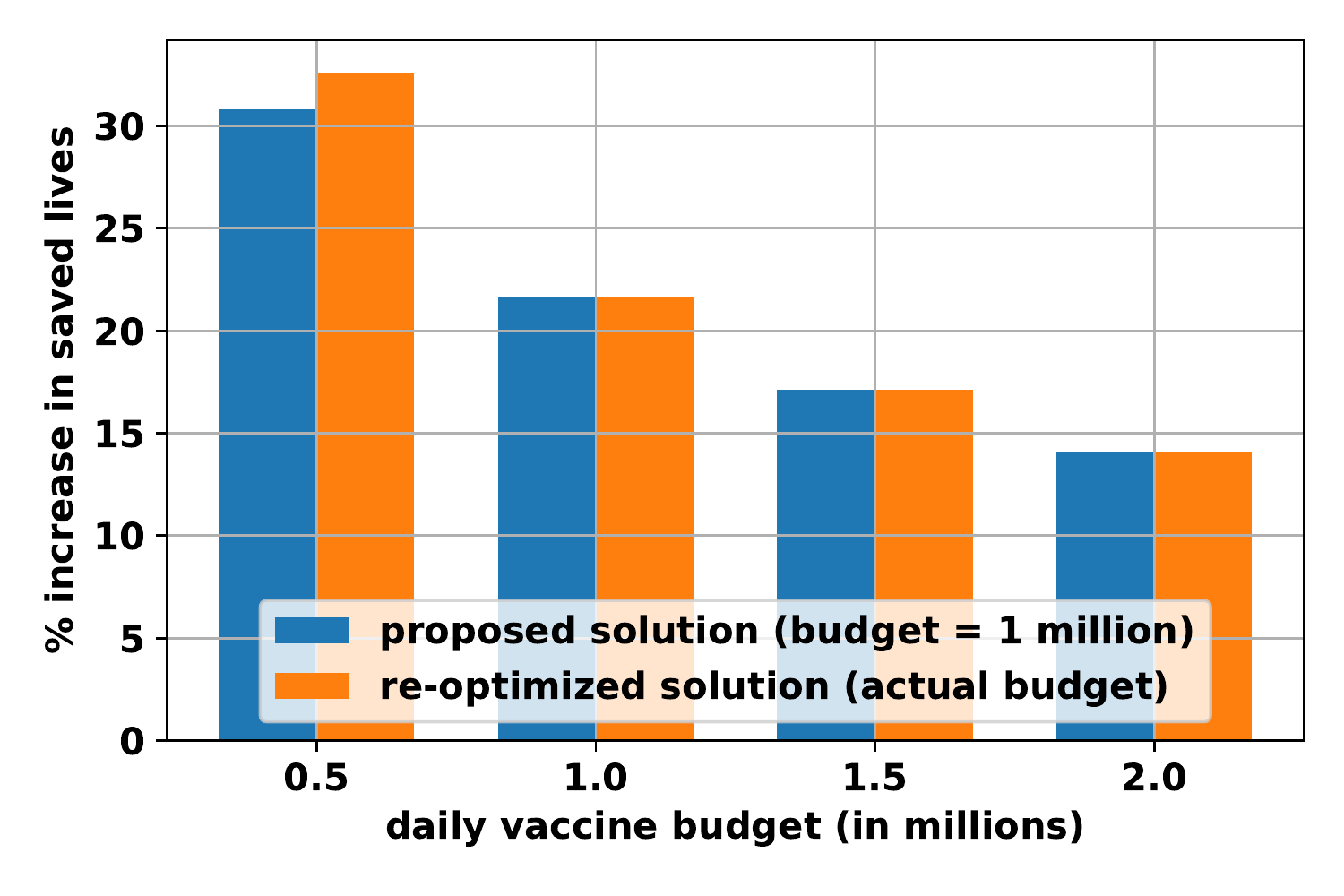}}
	\vspace{6 pt}
	\caption{Sensitivity and robustness of results with varying vaccine effectiveness and vaccine budget.}
	\label{fig:effectiveness_budget}
	\vspace{-12 pt}
\end{figure}

We first vary the two main drivers of the vaccination campaign, that are the vaccine effectiveness (parameter $\beta$) and the vaccine budget (parameters $B_t$). Figure~\ref{fig:effectiveness_budget} reports the percent-wise increase in saved lives, as compared to the top-cities benchmark, for both the proposed and the re-optimized solutions. The main takeaways fall under three categories. First, the benefits of optimization increase with vaccine effectiveness. This is expected, as a higher vaccine effectiveness increases the impact of all strategies (in the extreme example where $\beta=0$, all strategies have the same null performance). Second, the benefits of optimization decrease with vaccine budget. Indeed, in the extreme scenario with an infinite vaccine budget, any distribution strategy can immediately end the pandemic, leaving essentially no space for optimization. As the budget becomes more scarce, the decisions of \textit{who} receives a vaccine and \textit{when} become increasingly complex, so the optimized strategy has a positive and significant impact on the spread of the disease. It is interesting to note that this monotonic relationship would get inverted in the other regime with a small vaccine budget (again, all solutions perform identically when the budget gets to zero). Yet, with the current vaccine capacities, the benefits of optimization are very strong, saving an extra 15--35\% of lives.

The third takeaway from Figure~\ref{fig:effectiveness_budget} is the high degree of robustness of the proposed solution. In all but one experiment, the proposed solution remains optimal under the perturbed parameters (indicated by exactly the same benefits obtained with the proposed solution and the re-optimized solution). In the last case (with a daily budget of 500,000 vaccines), the proposed solution is dominated by the re-optimized solution, but remains within 2\% of the new optimum. Obviously, the new values of vaccine effectiveness and vaccine budgets impact downstream vaccine allocations and the dynamics of the pandemic. However, the location of vaccination sites is highly robust to variations in vaccine effectiveness and vaccine budgets.
%\footnote{The proposed and re-optimized solutions either result in the same set of vaccination sites or vary by one site. When evaluated with the full (bilinear) DELPHI--V model (as opposed to the approximate dynamics in the $\texttt{Optimize}$ module of our coordinate descent algorithm), the re-optimized solution results in a slightly higher death toll than the proposed solution. In other words, under each perturbation, we have not been able to find a solution that outperforms the proposed solution obtained prior to the perturbations.}

Next, we test the robustness of the proposed solution to the dynamics of the pandemic. One assumption of DELPHI--V is that the infection rate (captured by the parameters $\alpha_j$ and $\gamma_{jt}$) is identical across age groups. However, serological evidence suggests potential disparities in infections across sub-populations. To test this, we vary the infection rates with the risk class $k$, by adjusting the relative variations based on the serological estimates from the \citet{CDC}. Another assumption is that vaccines prevent mortality but not infection and transmission. In practice, vaccines may still reduce the risk of infection and the propensity to transmit the disease. To test this, we perturb the DELPHI--V model by assuming that a fraction $\beta$ of vaccinated people transition directly to the immune state $M$ (as opposed to the susceptible and vaccinated state $S'$).

\begin{figure}[h!]
    \centering
	\subfloat[Age-dependent infection rates]{\label{subfig:cdc_infection_rate}\includegraphics[width=.5\textwidth]{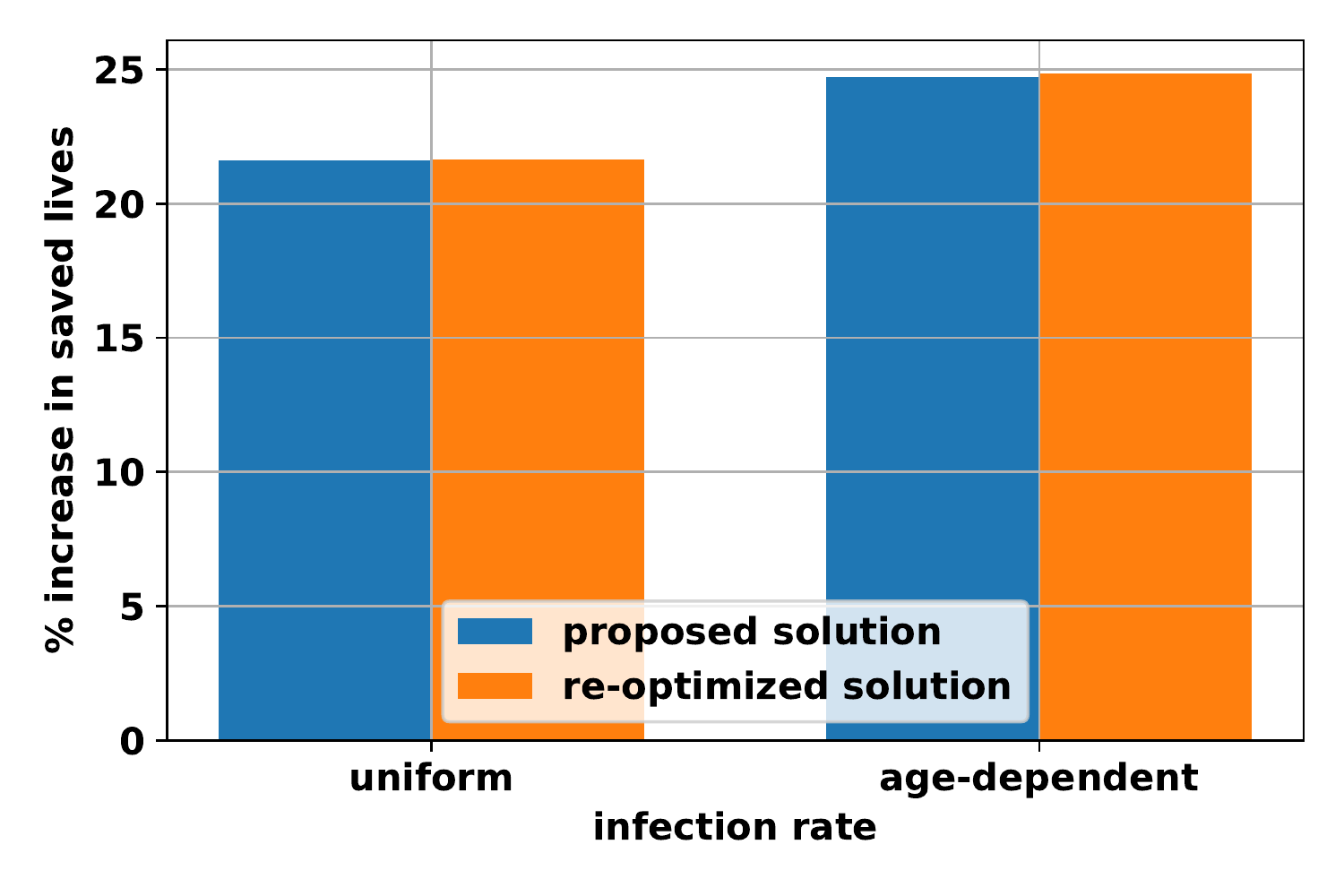}}
	\subfloat[No transmission]{\label{subfig:transmission}\includegraphics[width=.5\textwidth]{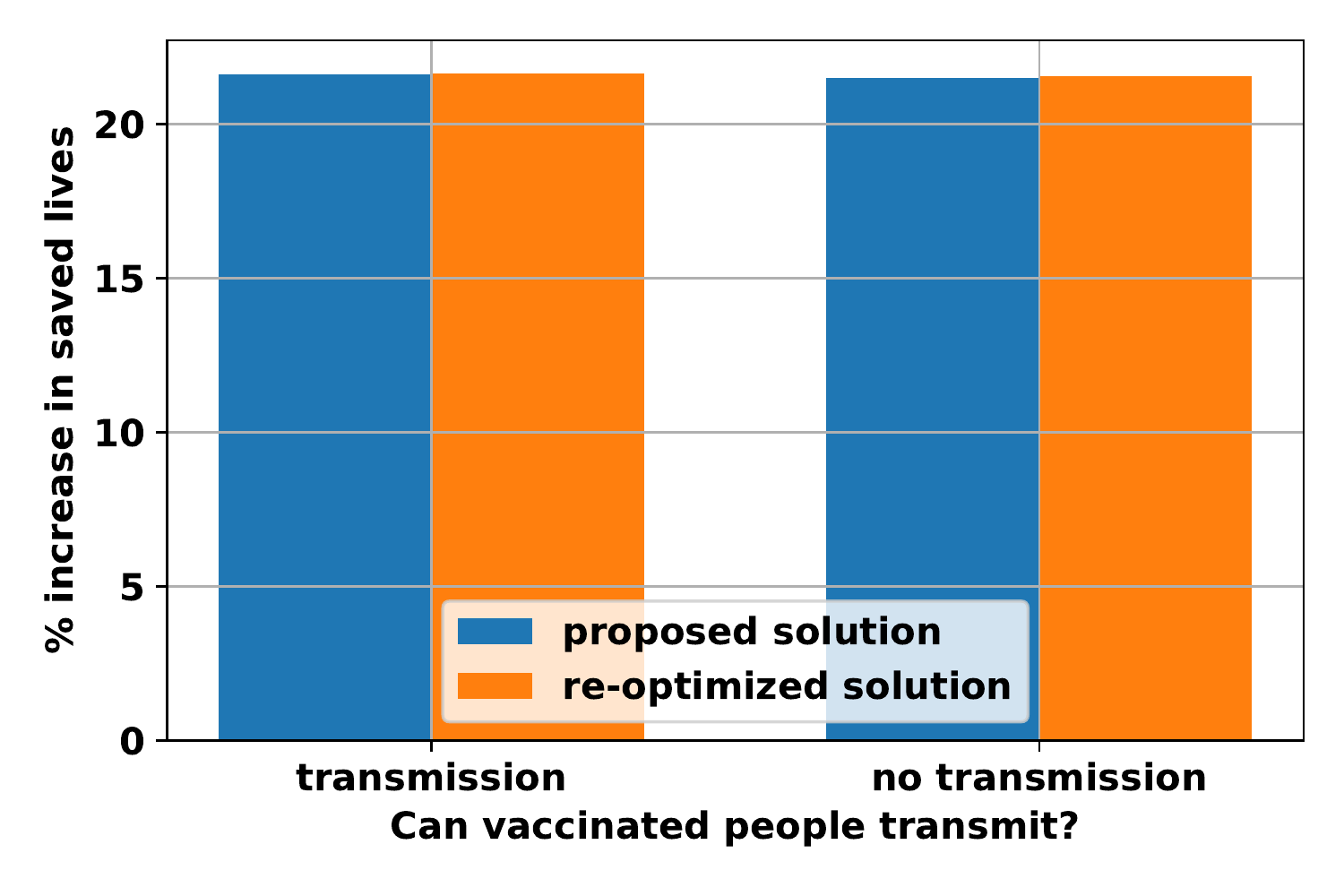}}
	\vspace{6 pt}
	\caption{Robustness of results with age-dependent infection rates and no transmission from vaccinated people.}
	\label{fig:more_robustness}
	\vspace{-12 pt}
\end{figure}

Figure~\ref{fig:more_robustness} reports the results from these new perturbations, using the same nomenclature as Figure~\ref{fig:effectiveness_budget}. The main takeaway is identical, in that the proposed solution remains near-optimal under the two perturbations. In both tests, the proposed solution results in less than 50 extra deaths than the re-optimized solution---a very small amount in view of the 20,000 lives saved by the optimization. In other words, the proposed solution is not only robust to vaccination characteristics (effectiveness and budget) but also to the dynamics of the COVID-19 pandemic.

Finally, we vary two key parameters in the DELPHI--V dynamics: the infection and mortality rates. Although these parameters are fitted from historical data, there remains considerable uncertainty regarding the dynamics of the pandemic and people's behaviors through a period of mass vaccination. Therefore, we introduce a random perturbation in the infection or mortality rate, in each state. Specifically, we define 50 perturbation scenarios; in each one, we sample each parameter in each state independently, following a uniform distribution centered around the nominal value and spanning $\pm20\%$. Therefore, the full range of infection and mortality rates spans 40\% of the nominal value---thus capturing instances where the estimated infection and mortality rates in the DELPHI--V model are subject to very large errors.

\begin{figure}[h!]
    \centering
	\subfloat[Infection rates]{\label{subfig:infection}\includegraphics[width=.5\textwidth]{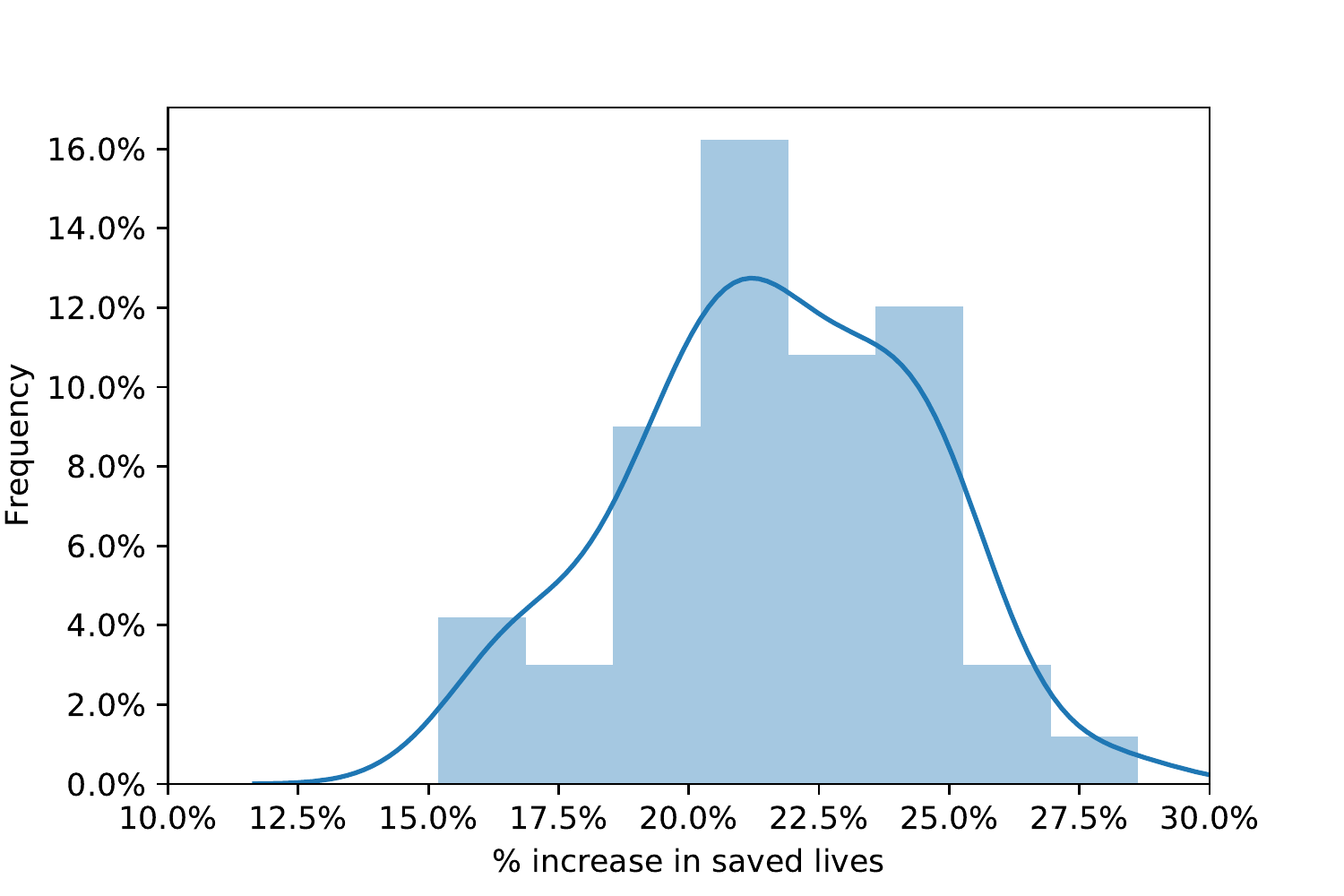}}
	\subfloat[Mortality rates]{\label{subfig:mortality}\includegraphics[width=.5\textwidth]{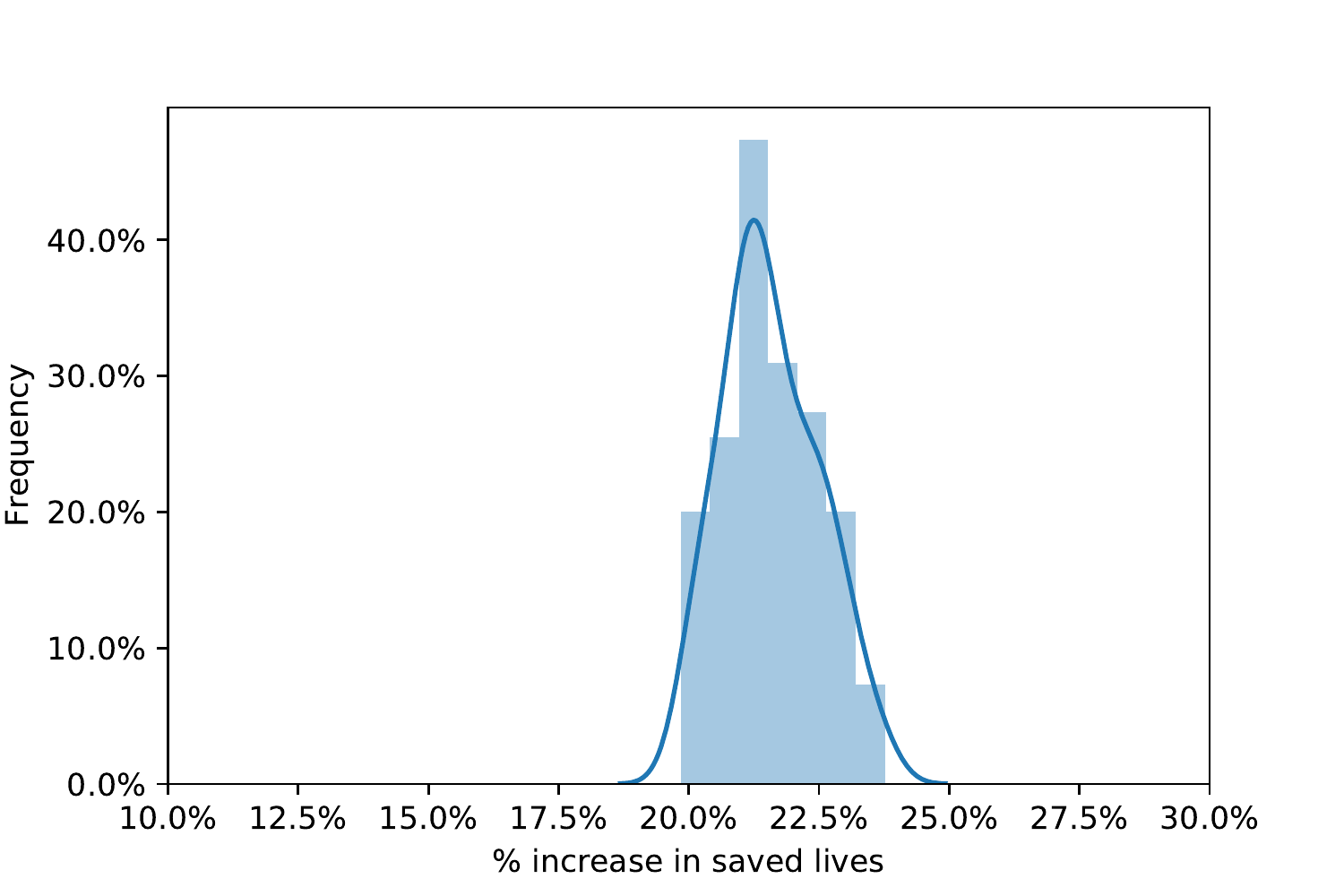}}
	\vspace{6 pt}
	\caption{Robustness of the benefits of optimization with infection and mortality rates.}
	\label{fig:infection_mortality}
	\vspace{-12 pt}
\end{figure}

Figure~\ref{fig:infection_mortality} reports the distribution of the benefits of optimization under these 50 scenarios, with perturbed infection rates (Figure~\ref{subfig:infection}) and mortality rates (Figure~\ref{subfig:mortality}). We compare here the proposed solution (obtained with the nominal values of the infection and mortality rates) and the top-cities benchmark, both evaluated with the perturbed infection and mortality rates. As the results show, the benefits of optimization remains highly significant, and robust to the perturbations. Recall that the benefits of optimization were estimated at 20\% under the nominal infection and mortality rates (Table~\ref{tab:saved}). After perturbations, they span 12.5\%--30\% with perturbations in the infection rates and 17.5\%--25\% with the perturbed mortality rates. In other words, even if the DELPHI--V model makes large errors when estimating the key dynamics of the pandemic, the proposed optimized vaccination sites still increase the impact of the vaccination campaign by over 10\%. In fact, the variations can go either way: in over half the simulations, the relative benefits of the optimized solution are even higher under the perturbed parameters than the nominal ones.
 
Note, finally, that the infection rate has a more significant impact on the relative benefits of optimized vaccine allocation than the mortality rate. This suggests that the impact of the pandemic depends mainly on how vaccines can curb infections at the upstream level---as opposed to mortality at the downstream level. This, again, illustrates the effects of the non-linear SEIR dynamics on the spread of the disease, and how we can leverage an epidemiological model such as DELPHI--V to allocate resources strategically in order to combat the pandemic most effectively.

From a practical standpoint, the results from this section are highly significant. Indeed, the DELPHI--V model, like any epidemiological model, only provides a rough approximation of the dynamics of the pandemic and the effects of vaccinations. Yet, the robustness of the solution provides guarantees that the locations of mass vaccination sites, even optimized against this approximate model, remain highly robust when evaluated against alternative dynamics. This is obviously not to say that we can commit to a full-scale vaccine distribution strategy that spans the entire vaccination campaign. However, the ``here-and-now'' decisions (i.e., the location of vaccination sites) are likely to remain near-optimal (or even optimal) in the next phases of the pandemic, ultimately enabling the vaccination campaign to save an extra 15--35\% of lives.

\section{Conclusion}
\label{conclusion}

This paper has presented a new prescriptive approach to optimize vaccine distribution strategies in response to the COVID-19 pandemic. The approach starts with a state-of-the-art epidemiological model called DELPHI, which augments SEIR models by capturing dynamics specific to COVID-19 (under-detection, governmental and societal response, and declining mortality rates). This paper has first proposed an extension, named DELPHI--V, which captures the effects of vaccinations and reflects the disaggregated impact of COVID-19 on mortality across risk classes (e.g., age groups). Then, this paper has embedded the predictive DELPHI--V model into an optimization model, termed DELPHI--V--OPT, to support vaccine distribution. DELPHI--V--OPT is formulated as a bilinear optimization model, and solved using a tailored algorithm based on coordinate descent.

We applied the model and algorithm to one of the priorities of the new Biden administration in the United States: determining the locations of mass vaccination sites across the country. We formalize the problem by selecting locations strategically to minimize the death toll of the pandemic, subject to practical and fairness constraints. Experimental results using real-world data suggest that the proposed optimization approach can yield significant benefits, as compared to benchmark solutions that locate vaccination sites based on readily-available demographic and epidemiological information. Specifically, the model can increase the effectiveness of the vaccination campaign by 20\%, saving an extra 4,000 lives over a three-month period. Remarkably, the proposed solution achieves critical fairness objectives---by significantly reducing the death toll of the pandemic in several states without hurting others---and is highly robust to uncertainties and forecast errors---by achieving similar benefits under a vast range of perturbations.

Obviously, the optimization approach developed in this paper is not without limitations. For instance, our experiments have only partitioned the population according to age groups, thus ignoring other objectives such as prioritizing allocations to healthcare workers, other essential workers, or patients with comorbidities. Moreover, our epidemiological model does not capture heterogeneity in population mixing across subpopulations (e.g., interactions may be more intense in urban areas and between young people than in rural areas and between older populations). Similarly, our model only captures the first-order effects of vaccinations, ignoring for instance heterogeneity across multiple vaccine types and double-dose vaccines. Finally, our methodology relies on a time discretization approximation and a coordinate descent approach, which do not yield theoretical guarantees on solution quality.

Whereas these limitations undoubtedly motivate further research, this paper lays one of the first data-driven bricks on the optimal distribution of COVID-19 vaccines at a macroscopic level. At a time where vaccine development and vaccine production are going full speed, the results from this paper highlight the critical role of vaccine distribution strategies to combat the pandemic. Obviously, it is essential to make every effort possible to develop new vaccines, enhance vaccine effectiveness, and produce as many vaccines as possible. But this paper highlights another lever that can be pulled to curb the effect of the pandemic: strategically managing vaccine stockpiles to prevent the spread of the pandemic and mitigate its impact. As such, this paper can provide critical decision-making support to governmental agencies as they are currently planning mass vaccinations around the globe.

\section*{Acknowledgments}
The authors gratefully acknowledge T. Greg McKelvey Jr. for feedback and discussions that have greatly improved the quality of the paper.
	
\bibliographystyle{informs2014}
\bibliography{bibliography}

\begin{thebibliography}{63}
\providecommand{\natexlab}[1]{#1}
\providecommand{\url}[1]{\texttt{#1}}
\providecommand{\urlprefix}{URL }

\bibitem[{Aaby et~al.(2006)Aaby, Herrmann, Jordan, Treadwell, \protect\BIBand{}
  Wood}]{aaby2006montgomery}
Aaby K, Herrmann JW, Jordan CS, Treadwell M, Wood K (2006) {Montgomery
  county’s public health service uses operations research to plan emergency
  mass dispensing and vaccination clinics}. \emph{Interfaces} 36(6):569--579.

\bibitem[{Araz et~al.(2012)Araz, Galvani, \protect\BIBand{}
  Meyers}]{araz2012geographic}
Araz OM, Galvani A, Meyers LA (2012) {Geographic prioritization of distributing
  pandemic influenza vaccines}. \emph{Health Care Management Science}
  15(3):175--187.

\bibitem[{Arifo{\u{g}}lu et~al.(2012)Arifo{\u{g}}lu, Deo, \protect\BIBand{}
  Iravani}]{arifouglu2012consumption}
Arifo{\u{g}}lu K, Deo S, Iravani SM (2012) {Consumption externality and yield
  uncertainty in the influenza vaccine supply chain: Interventions in demand
  and supply sides}. \emph{Management Science} 58(6):1072--1091.

\bibitem[{Arinaminpathy et~al.(2012)Arinaminpathy, Ratmann, Koelle, Epstein,
  Price, Viboud, Miller, \protect\BIBand{} Grenfell}]{arinaminpathy2012impact}
Arinaminpathy N, Ratmann O, Koelle K, Epstein SL, Price GE, Viboud C, Miller
  MA, Grenfell BT (2012) Impact of cross-protective vaccines on epidemiological
  and evolutionary dynamics of influenza. \emph{Proceedings of the National
  Academy of Sciences} 109(8):3173--3177.

\bibitem[{Bae et~al.(2020)Bae, Gandhi, Kothari, Shankar, Bae, Patwa, Sukumaran,
  Mishra, Murali, Saxena et~al.}]{bae2020challenges}
Bae J, Gandhi D, Kothari J, Shankar S, Bae J, Patwa P, Sukumaran R, Mishra K,
  Murali S, Saxena A, et~al. (2020) Challenges of equitable vaccine
  distribution in the covid-19 pandemic. \emph{arXiv preprint arXiv:2012.12263}
  .

\bibitem[{Bandi \protect\BIBand{} Bertsimas(2020)}]{bandi2020optimizing}
Bandi H, Bertsimas D (2020) Optimizing influenza vaccine composition: From
  predictions to prescriptions. \emph{Machine Learning for Healthcare
  Conference}, 121--142 (PMLR).

\bibitem[{Basta et~al.(2009)Basta, Chao, Halloran, Matrajt, \protect\BIBand{}
  Longini~Jr}]{basta2009strategies}
Basta NE, Chao DL, Halloran ME, Matrajt L, Longini~Jr IM (2009) {Strategies for
  pandemic and seasonal influenza vaccination of schoolchildren in the United
  States}. \emph{American journal of epidemiology} 170(6):679--686.

\bibitem[{Bertsimas et~al.(2020)Bertsimas, Boussioux, Wright, Delarue,
  Digalakis~Jr, Jacquillat, Kitane, Lukin, Li, Mingardi et~al.}]{bertsimas:20}
Bertsimas D, Boussioux L, Wright RC, Delarue A, Digalakis~Jr V, Jacquillat A,
  Kitane DL, Lukin G, Li ML, Mingardi L, et~al. (2020) {From predictions to
  prescriptions: A data-driven response to COVID-19}. \emph{arXiv preprint
  arXiv:2006.16509} .

\bibitem[{Chick et~al.(2008)Chick, Mamani, \protect\BIBand{}
  Simchi-Levi}]{chick2008supply}
Chick SE, Mamani H, Simchi-Levi D (2008) Supply chain coordination and
  influenza vaccination. \emph{Operations Research} 56(6):1493--1506.

\bibitem[{Cho(2010)}]{cho2010optimal}
Cho SH (2010) The optimal composition of influenza vaccines subject to random
  production yields. \emph{Manufacturing \& Service Operations Management}
  12(2):256--277.

\bibitem[{Chowell et~al.(2009)Chowell, Viboud, Wang, Bertozzi,
  \protect\BIBand{} Miller}]{chowell2009adaptive}
Chowell G, Viboud C, Wang X, Bertozzi SM, Miller MA (2009) Adaptive vaccination
  strategies to mitigate pandemic influenza: Mexico as a case study. \emph{PLoS
  One} 4(12):e8164.

\bibitem[{Coustasse et~al.(2021)Coustasse, Kimble, \protect\BIBand{}
  Maxik}]{coustasse2021covid}
Coustasse A, Kimble C, Maxik K (2021) Covid-19 and vaccine hesitancy: A
  challenge the united states must overcome. \emph{The Journal of Ambulatory
  Care Management} 44(1):71--75.

\bibitem[{Dai et~al.(2016)Dai, Cho, \protect\BIBand{}
  Zhang}]{dai2016contracting}
Dai T, Cho SH, Zhang F (2016) Contracting for on-time delivery in the us
  influenza vaccine supply chain. \emph{Manufacturing \& Service Operations
  Management} 18(3):332--346.

\bibitem[{Dror et~al.(2020)Dror, Eisenbach, Taiber, Morozov, Mizrachi, Zigron,
  Srouji, \protect\BIBand{} Sela}]{dror2020vaccine}
Dror AA, Eisenbach N, Taiber S, Morozov NG, Mizrachi M, Zigron A, Srouji S,
  Sela E (2020) Vaccine hesitancy: the next challenge in the fight against
  covid-19. \emph{European journal of epidemiology} 35(8):775--779.

\bibitem[{Duijzer et~al.(2018)Duijzer, van Jaarsveld, \protect\BIBand{}
  Dekker}]{duijzer2018literature}
Duijzer LE, van Jaarsveld W, Dekker R (2018) Literature review: The vaccine
  supply chain. \emph{European Journal of Operational Research}
  268(1):174--192.

\bibitem[{Dushoff et~al.(2007)Dushoff, Plotkin, Viboud, Simonsen, Miller, Loeb,
  \protect\BIBand{} David}]{dushoff2007vaccinating}
Dushoff J, Plotkin JB, Viboud C, Simonsen L, Miller M, Loeb M, David J (2007)
  Vaccinating to protect a vulnerable subpopulation. \emph{PLoS Med} 4(5):e174.

\bibitem[{Elveback et~al.(1976)Elveback, Fox, Ackerman, Langworthy, Boyd,
  \protect\BIBand{} Gatewood}]{elveback:76}
Elveback LR, Fox JP, Ackerman E, Langworthy A, Boyd M, Gatewood L (1976) An
  influenza simulation model for immunization studies. \emph{American Journal
  of Epidemiology} 103(2):152--165.

\bibitem[{Emanuel \protect\BIBand{} Wertheimer(2006)}]{emanuel2006should}
Emanuel EJ, Wertheimer A (2006) Who should get influenza vaccine when not all
  can? \emph{Science} 312(5775):854--855.

\bibitem[{{Federal Emergency Management
  Agency}(2021{\natexlab{a}})}]{fema2021announcement2}
{Federal Emergency Management Agency} (2021{\natexlab{a}}) {FEMA COVID-19
  Vaccination Update}.
  \url{https://www.fema.gov/press-release/20210402/fema-covid-19-vaccination-update},
  accessed: 2021-04-09.

\bibitem[{{Federal Emergency Management
  Agency}(2021{\natexlab{b}})}]{fema2021announcement1}
{Federal Emergency Management Agency} (2021{\natexlab{b}}) {FEMA Supporting
  Vaccination Centers Nationwide}.
  \url{https://www.fema.gov/press-release/20210226/fema-supporting-vaccination-centers-nationwide},
  accessed: 2021-04-09.

\bibitem[{Federgruen \protect\BIBand{} Yang(2009)}]{federgruen2009competition}
Federgruen A, Yang N (2009) Competition under generalized attraction models:
  Applications to quality competition under yield uncertainty. \emph{Management
  science} 55(12):2028--2043.

\bibitem[{Florindo et~al.(2020)Florindo, Kleiner, Vaskovich-Koubi, Ac{\'u}rcio,
  Carreira, Yeini, Tiram, Liubomirski, \protect\BIBand{}
  Satchi-Fainaro}]{florindo2020immune}
Florindo HF, Kleiner R, Vaskovich-Koubi D, Ac{\'u}rcio RC, Carreira B, Yeini E,
  Tiram G, Liubomirski Y, Satchi-Fainaro R (2020) {Immune-mediated approaches
  against COVID-19}. \emph{Nature nanotechnology} 1--16.

\bibitem[{Goyal et~al.(2020)Goyal, Choi, Pinheiro, Schenck, Chen, Jabri,
  Satlin, Campion~Jr, Nahid, Ringel et~al.}]{goyal2020clinical}
Goyal P, Choi JJ, Pinheiro LC, Schenck EJ, Chen R, Jabri A, Satlin MJ,
  Campion~Jr TR, Nahid M, Ringel JB, et~al. (2020) {Clinical characteristics of
  Covid-19 in New York City}. \emph{New England Journal of Medicine} .

\bibitem[{Graham(2020)}]{graham2020rapid}
Graham BS (2020) {Rapid COVID-19 vaccine development}. \emph{Science}
  368(6494):945--946.

\bibitem[{Grauer et~al.(2020)Grauer, L{\"o}wen, \protect\BIBand{}
  Liebchen}]{grauer2020strategic}
Grauer J, L{\"o}wen H, Liebchen B (2020) Strategic spatiotemporal vaccine
  distribution increases the survival rate in an infectious disease like
  covid-19. \emph{Scientific reports} 10(1):1--10.

\bibitem[{Guan et~al.(2020)Guan, Ni, Hu, Liang, Ou, He, Liu, Shan, Lei, Hui
  et~al.}]{guan2020clinical}
Guan Wj, Ni Zy, Hu Y, Liang Wh, Ou Cq, He Jx, Liu L, Shan H, Lei Cl, Hui DS,
  et~al. (2020) {Clinical characteristics of coronavirus disease 2019 in
  China}. \emph{New England journal of medicine} 382(18):1708--1720.

\bibitem[{{Gurobi Optimization}(2020)}]{gurobi}
{Gurobi Optimization} (2020) Gurobi optimizer reference manual.
  \urlprefix\url{http://www.gurobi.com}.

\bibitem[{Jacobson et~al.(2006{\natexlab{a}})Jacobson, Sewell,
  \protect\BIBand{} Proano}]{jacobson2006analysis}
Jacobson SH, Sewell EC, Proano RA (2006{\natexlab{a}}) An analysis of the
  pediatric vaccine supply shortage problem. \emph{Health Care Management
  Science} 9(4):371--389.

\bibitem[{Jacobson et~al.(2006{\natexlab{b}})Jacobson, Sewell, Proano,
  \protect\BIBand{} Jokela}]{jacobson2006stockpile}
Jacobson SH, Sewell EC, Proano RA, Jokela JA (2006{\natexlab{b}}) Stockpile
  levels for pediatric vaccines: How much is enough? \emph{Vaccine}
  24(17):3530--3537.

\bibitem[{Keeling \protect\BIBand{} Shattock(2012)}]{keeling2012optimal}
Keeling MJ, Shattock A (2012) Optimal but unequitable prophylactic distribution
  of vaccine. \emph{Epidemics} 4(2):78--85.

\bibitem[{Khamsi(2020)}]{khamsi2020if}
Khamsi R (2020) If a coronavirus vaccine arrives, can the world make enough.
  \emph{Nature} 580(7805):578--580.

\bibitem[{Kissler et~al.(2020)Kissler, Tedijanto, Goldstein, Grad,
  \protect\BIBand{} Lipsitch}]{kissler:20}
Kissler SM, Tedijanto C, Goldstein E, Grad YH, Lipsitch M (2020) {Projecting
  the transmission dynamics of SARS-CoV-2 through the postpandemic period}.
  \emph{Science} .

\bibitem[{Kornish \protect\BIBand{} Keeney(2008)}]{kornish2008repeated}
Kornish LJ, Keeney RL (2008) Repeated commit-or-defer decisions with a
  deadline: The influenza vaccine composition. \emph{Operations Research}
  56(3):527--541.

\bibitem[{Lee et~al.(2012)Lee, Golinski, \protect\BIBand{}
  Chowell}]{lee2012modeling}
Lee S, Golinski M, Chowell G (2012) Modeling optimal age-specific vaccination
  strategies against pandemic influenza. \emph{Bulletin of mathematical
  biology} 74(4):958--980.

\bibitem[{Lemmens et~al.(2016)Lemmens, Decouttere, Vandaele, \protect\BIBand{}
  Bernuzzi}]{lemmens2016review}
Lemmens S, Decouttere C, Vandaele N, Bernuzzi M (2016) A review of integrated
  supply chain network design models: Key issues for vaccine supply chains.
  \emph{Chemical Engineering Research and Design} 109:366--384.

\bibitem[{Li et~al.(2020)Li, Tazi~Bouardi, Skali~Lami, Trikalinos, Trichakis,
  \protect\BIBand{} Bertsimas}]{li:20}
Li ML, Tazi~Bouardi H, Skali~Lami O, Trikalinos TA, Trichakis NK, Bertsimas D
  (2020) {Forecasting COVID-19 and analyzing the effect of government
  interventions}. \emph{medRxiv}
  \urlprefix\url{https://www.medrxiv.org/content/early/2020/06/24/2020.06.23.20138693}.

\bibitem[{Longini~Jr et~al.(1978)Longini~Jr, Ackerman, \protect\BIBand{}
  Elveback}]{longini:78}
Longini~Jr IM, Ackerman E, Elveback LR (1978) An optimization model for
  influenza a epidemics. \emph{Mathematical Biosciences} 38(1-2):141--157.

\bibitem[{Lurie et~al.(2020)Lurie, Saville, Hatchett, \protect\BIBand{}
  Halton}]{lurie2020developing}
Lurie N, Saville M, Hatchett R, Halton J (2020) Developing covid-19 vaccines at
  pandemic speed. \emph{New England Journal of Medicine} 382(21):1969--1973.

\bibitem[{Mak et~al.(2021)Mak, Dai, \protect\BIBand{} Tang}]{mak2021managing}
Mak HY, Dai T, Tang CS (2021) Managing two-dose covid-19 vaccine rollouts with
  limited supply. \emph{Available at SSRN 3790836} .

\bibitem[{Mamani et~al.(2013)Mamani, Chick, \protect\BIBand{}
  Simchi-Levi}]{mamani2013game}
Mamani H, Chick SE, Simchi-Levi D (2013) A game-theoretic model of
  international influenza vaccination coordination. \emph{Management Science}
  59(7):1650--1670.

\bibitem[{Matrajt et~al.(2020)Matrajt, Eaton, Leung, \protect\BIBand{}
  Brown}]{matrajt2020vaccine}
Matrajt L, Eaton J, Leung T, Brown ER (2020) Vaccine optimization for covid-19:
  who to vaccinate first? \emph{medRxiv} .

\bibitem[{Matrajt et~al.(2013)Matrajt, Halloran, \protect\BIBand{}
  Longini~Jr}]{matrajt2013optimal}
Matrajt L, Halloran ME, Longini~Jr IM (2013) Optimal vaccine allocation for the
  early mitigation of pandemic influenza. \emph{PLoS Comput Biol}
  9(3):e1002964.

\bibitem[{Medlock \protect\BIBand{} Galvani(2009)}]{medlock2009optimizing}
Medlock J, Galvani AP (2009) Optimizing influenza vaccine distribution.
  \emph{Science} 325(5948):1705--1708.

\bibitem[{Muriel \protect\BIBand{} Bauchner(2021)}]{muriel2021vaccine}
Muriel JJ, Bauchner H (2021) Vaccine distribution—equity left behind?
  \emph{JAMA} .

\bibitem[{{National Academies of Sciences, Engineering, and
  Medicine}(2020)}]{national2020framework}
{National Academies of Sciences, Engineering, and Medicine} (2020) {Framework
  for equitable allocation of COVID-19 vaccine} .

\bibitem[{{New York Times}(2020)}]{nyt:20}
{New York Times} (2020) {New York Times, Coronavirus in the U.S.: Latest Map
  and Case Count}.
  {https://www.nytimes.com/interactive/2020/us/coronavirus-us-cases.html}.

\bibitem[{Patel et~al.(2005)Patel, Longini~Jr, \protect\BIBand{}
  Halloran}]{patel2005finding}
Patel R, Longini~Jr IM, Halloran ME (2005) Finding optimal vaccination
  strategies for pandemic influenza using genetic algorithms. \emph{Journal of
  theoretical biology} 234(2):201--212.

\bibitem[{Perkins \protect\BIBand{} Espana(2020)}]{notredame}
Perkins A, Espana G (2020) Notredame-fred covid-19 forecasts.

\bibitem[{Petrilli et~al.(2020)Petrilli, Jones, Yang, Rajagopalan, O'Donnell,
  Chernyak, Tobin, Cerfolio, Francois, \protect\BIBand{}
  Horwitz}]{petrilli2020factors}
Petrilli CM, Jones SA, Yang J, Rajagopalan H, O'Donnell LF, Chernyak Y, Tobin
  K, Cerfolio RJ, Francois F, Horwitz LI (2020) {Factors associated with
  hospitalization and critical illness among 4,103 patients with COVID-19
  disease in New York City}. \emph{medRxiv} .

\bibitem[{Rastegar et~al.(2021)Rastegar, Tavana, Meraj, \protect\BIBand{}
  Mina}]{rastegar2021inventory}
Rastegar M, Tavana M, Meraj A, Mina H (2021) An inventory-location optimization
  model for equitable influenza vaccine distribution in developing countries
  during the covid-19 pandemic. \emph{Vaccine} 39(3):495--504.

\bibitem[{Rodriguez et~al.(2020)Rodriguez, Tabassum, Cui, Xie, Ho, Agarwal,
  Adhikari, \protect\BIBand{} Prakash}]{rodriguez2020deepcovid}
Rodriguez A, Tabassum A, Cui J, Xie J, Ho J, Agarwal P, Adhikari B, Prakash BA
  (2020) Deepcovid: An operational deep learning-driven framework for
  explainable real-time covid-19 forecasting. \emph{medRxiv} .

\bibitem[{Shin et~al.(2020)Shin, Shukla, Chung, Beiss, Chan, Ortega-Rivera,
  Wirth, Chen, Sack, Pokorski et~al.}]{shin2020covid}
Shin MD, Shukla S, Chung YH, Beiss V, Chan SK, Ortega-Rivera OA, Wirth DM, Chen
  A, Sack M, Pokorski JK, et~al. (2020) {COVID-19 vaccine development and a
  potential nanomaterial path forward}. \emph{Nature Nanotechnology} 1--10.

\bibitem[{Sun et~al.(2009)Sun, Yang, \protect\BIBand{}
  De~V{\'e}ricourt}]{sun2009selfish}
Sun P, Yang L, De~V{\'e}ricourt F (2009) Selfish drug allocation for containing
  an international influenza pandemic at the onset. \emph{Operations Research}
  57(6):1320--1332.

\bibitem[{Tanner et~al.(2008)Tanner, Sattenspiel, \protect\BIBand{}
  Ntaimo}]{tanner2008finding}
Tanner MW, Sattenspiel L, Ntaimo L (2008) Finding optimal vaccination
  strategies under parameter uncertainty using stochastic programming.
  \emph{Mathematical biosciences} 215(2):144--151.

\bibitem[{Teytelman \protect\BIBand{}
  Larson(2013)}]{teytelman2013multiregional}
Teytelman A, Larson RC (2013) Multiregional dynamic vaccine allocation during
  an influenza epidemic. \emph{Service Science} 5(3):197--215.

\bibitem[{Uribe-S{\'a}nchez et~al.(2011)Uribe-S{\'a}nchez, Savachkin, Santana,
  Prieto-Santa, \protect\BIBand{} Das}]{uribe2011predictive}
Uribe-S{\'a}nchez A, Savachkin A, Santana A, Prieto-Santa D, Das TK (2011) A
  predictive decision-aid methodology for dynamic mitigation of influenza
  pandemics. \emph{OR spectrum} 33(3):751--786.

\bibitem[{{US Census Bureau}(2020)}]{uscensus:01}
{US Census Bureau} (2020) {County Population by Characteristics: 2010-2019}.
  https://www.census.gov/data/tables/time-series/demo/popest/2010s-counties-detail.html.

\bibitem[{{US Center for Disease Control}(2020{\natexlab{a}})}]{CDC}
{US Center for Disease Control} (2020{\natexlab{a}}) {COVID-19 Forecasts}.
  {https://www.cdc.gov/coronavirus/2019-ncov/covid-data/forecasting-us.html}.

\bibitem[{{US Center for Disease Control}(2020{\natexlab{b}})}]{cdc:death_mult}
{US Center for Disease Control} (2020{\natexlab{b}}) {COVID-19 Hospitalization
  and Death by Age}.
  https://www.cdc.gov/coronavirus/2019-ncov/covid-data/investigations-discovery/hospitalization-death-by-age.html.

\bibitem[{{US Center for Disease Control}(2021)}]{cdc:death_count}
{US Center for Disease Control} (2021) {Provisional Death Counts for
  Coronavirus Disease 2019 (COVID-19)}.
  https://www.cdc.gov/nchs/nvss/vsrr/covid\_weekly.

\bibitem[{Watson(1972)}]{watson:72}
Watson RK (1972) On an epidemic in a stratified population. \emph{Journal of
  Applied Probability} 9(3):659--666.

\bibitem[{Wu et~al.(2005)Wu, Wein, \protect\BIBand{}
  Perelson}]{wu2005optimization}
Wu JT, Wein LM, Perelson AS (2005) Optimization of influenza vaccine selection.
  \emph{Operations Research} 53(3):456--476.

\bibitem[{Yarmand et~al.(2014)Yarmand, Ivy, Denton, \protect\BIBand{}
  Lloyd}]{yarmand2014optimal}
Yarmand H, Ivy JS, Denton B, Lloyd AL (2014) Optimal two-phase vaccine
  allocation to geographically different regions under uncertainty.
  \emph{European Journal of Operational Research} 233(1):208--219.

\end{thebibliography}

\section*{Appendix}

Table \ref{tab:proposed_centers} presents the proposed center allocation; for each state, we show the number of centers allocated, the number of vaccines per day, and the cities where the proposed centers are located.

\begin{table}[h]
\centering
\resizebox{\textwidth}{!}{
\begin{tabular}{|c|c|c|l|}
\hline
\textbf{State} & \textbf{\# centers} & \textbf{\# vaccines/day} & \textbf{Selected cities} \\ \hline
Alabama & 2 & 14458 & Birmingham, Mobile \\
Alaska & 1 & 6592 & Anchorage \\
Arizona & 2 & 29663 & Mesa, Scottsdale \\
Arkansas & 1 & 14094 & Little Rock \\
California & 8 & 52734 & Bakersfield, Chula Vista, Citrus Heights, Fremont, Fresno \\
 & & & Glendale, Indio, Irvine \\
Colorado & 2 & 28289 & Centennial, Colorado Springs \\
Connecticut & 1 & 8699 & Waterbury \\
Delaware & 1 & 6592 & Wilmington \\
District of Columbia & 1 & 6592 & Washington \\
Florida & 6 & 88989 & Boca Raton, Jacksonville, Miami, Orlando, Tallahassee, Tampa \\
Georgia & 2 & 13357 & Sandy Springs, Warner Robins \\
Hawaii & 1 & 6592 & Honolulu \\
Idaho & 1 & 6592 & Meridian \\
Illinois & 2 & 13184 & Cicero, Springfield \\
Indiana & 2 & 20807 & Indianapolis, South Bend \\
Iowa & 2 & 16356 & Cedar Rapids, Des Moines \\
Kansas & 1 & 6592 & Topeka \\
Kentucky & 1 & 6592 & Louisville \\
Louisiana & 2 & 29386 & Baton Rouge, Shreveport \\
Maine & 1 & 6592 & Portland \\
Maryland & 1 & 12167 & Baltimore \\
Massachusetts & 1 & 11756 & Lowell \\
Michigan & 2 & 13184 & Farmington Hills, Wyoming \\
Minnesota & 1 & 6592 & Brooklyn Park \\
Mississippi & 1 & 8991 & Jackson \\
Missouri & 2 & 19139 & Lee's Summit, O'Fallon \\
Montana & 1 & 6592 & Missoula \\
Nebraska & 1 & 6592 & Omaha \\
Nevada & 2 & 13184 & Henderson, Sparks \\
New Hampshire & 1 & 6592 & Nashua \\
New Jersey & 2 & 29663 & Camden, Newark \\
New Mexico & 1 & 11670 & Rio Rancho \\
New York & 5 & 70791 & Buffalo, Rochester, Schenectady, Syracuse, Yonkers \\
North Carolina & 2 & 13184 & Cary, Charlotte \\
North Dakota & 1 & 6592 & Fargo \\
Ohio & 3 & 19775 & Cincinnati, Cleveland, Columbus \\
Oklahoma & 2 & 20835 & Broken Arrow, Edmond \\
Oregon & 1 & 13395 & Portland \\
Pennsylvania & 2 & 13184 & Pittsburgh, Reading \\
Rhode Island & 1 & 6592 & Pawtucket \\
South Carolina & 1 & 10979 & Columbia \\
South Dakota & 1 & 6592 & Sioux Falls \\
Tennessee & 3 & 19775 & Knoxville, Memphis, Nashville \\
Texas & 11 & 163146 & Austin, Corpus Christi, Dallas, El Paso, Fort Worth, Houston, \\
 & & & Lubbock, McAllen, Midland, San Antonio, Tyler \\
Utah & 1 & 10818 & West Valley City \\
Vermont & 1 & 6804 & Burlington \\
Virginia & 3 & 22841 & Alexandria, Newport News, Roanoke \\
Washington & 2 & 26447 & Kent, Spokane \\
West Virginia & 1 & 8370 & Charleston \\
Wisconsin & 2 & 13184 & Appleton, Waukesha \\
Wyoming & 1 & 6592 & Cheyenne \\ \hline
\end{tabular}
}
\caption{Proposed center allocation}
\label{tab:proposed_centers}
\end{table}

{\color{black}
In Table \ref{tab:realized_centers}, we outline the realized center allocations which has so far been implemented by the Federal Emergency Management Agency (FEMA) \citep{fema2021announcement1,fema2021announcement2}. In Table \ref{tab:comparison_centers}, we compare the realized center allocation (R) with the proposed center allocation (Pr). For the latter, we calculate percentages by taking into account only the 48 centers that were allocated to the 13 states under consideration. We also show the population-based center allocation (Pop), for which we calculate percentages by taking into account the 54 centers that were allocated to the 13 states under consideration. The last column in Table \ref{tab:comparison_centers} indicates whether (R) and (Pr) agree, that is, whether the fraction of centers allocated by both to any given state is greater/equal/less than the population share of the state.

Similar to Table \ref{tab:comparison_centers}, Table \ref{tab:comparison_vaccines} compares the number of vaccines per day (v/d, in thousands) allocated to each state in the realized solution (R) with that in the proposed solution (Pr) and the population-based baseline solution (Pop). For (Pop), we take the (Pop) center allocation shown in Table \ref{tab:comparison_centers} and assume that the vaccines are distributed uniformly across centers. More specifically, the number of vaccines that each center gets is equal to the average number of vaccines allocated to each center in (R), namely, in the realized solution.

Both tables show that, despite obvious differences given the rollout plan and a number of practical considerations, the model-based recommendations are generally in line with the center and vaccine allocation decisions that were made in practice. Specifically, the realized and proposed solutions agree in the majority of cases, both in terms of centers and in terms of vaccines. These results underscore the role of our model as a strategic tool to support the deployment of vaccination sites as well as a tactical tool to support vaccine distribution.

}

\begin{table}[h]
{\color{black}
\centering
\resizebox{\textwidth}{!}{
\begin{tabular}{cccc}
\hline
\textbf{State} & \textbf{\# centers} & \textbf{\# vaccines/day} & \textbf{Selected cities} \\ \hline
California & 2 & 12000 & Oakland, Los Angeles \\
Florida & 4 & 12000 & Miami, Tampa, Orlando, Jacksonville \\
Illinois & 1 & 6000 & Chicago \\
Indiana & 1 & 3000 & Gary \\
Massachusetts & 1 & 5000 & Boston \\
Missouri & 1 & 3000 & St. Louis \\
New Jersey & 1 & 5000 & Newark \\
New York & 6 & 12000 & Brooklyn, Queens, Yonkers, Rochester, Buffalo, Albany \\
North Carolina & 1 & 3000 & Greensboro \\
Pennsylvania & 1 & 6000 & Philadelphia, \\
Texas & 3 & 12000 & Houston, Dallas, Arlington \\
Virginia & 1 & 5000 & Norfolk \\
Washington & 1 & 1000 & Yakima \\ \hline
\end{tabular}
}
\caption{Realized center allocation}
\label{tab:realized_centers}
}
\end{table}

\begin{table}[h]
{\color{black}
\centering
\footnotesize{
\begin{tabular}{l cc c cc c cc c}
\hline
    &   \multicolumn{2}{c}{Realized}    &&  \multicolumn{2}{c}{Proposed}    &&  \multicolumn{2}{c}{Population}   &   \\ \cmidrule{2-3}\cmidrule{5-6}\cmidrule{8-9}
\textbf{State} & \textbf{\#} & \textbf{\%} && \textbf{\#} & \textbf{\%} && \textbf{\#} & \textbf{\%} & \textbf{(R)$\iff$(Pr)} \\ \hline
California & 2 & 0.08 && 8 & 0.17 && 12 & 0.22 & 1 \\ \hline
Florida & 4 & 0.17 && 6 & 0.13 && 6 & 0.11 & 1 \\ \hline
Illinois & 1 & 0.04 && 2 & 0.04 && 4 & 0.07 & 1 \\ \hline
Indiana & 1 & 0.04 && 2 & 0.04 && 2 & 0.04 & 1 \\ \hline
Massachusetts & 1 & 0.04 && 1 & 0.02 && 2 & 0.04 & 0 \\ \hline
Missouri & 1 & 0.04 && 2 & 0.04 && 1 & 0.02 & 1 \\ \hline
New Jersey & 1 & 0.04 && 2 & 0.04 && 2 & 0.04 & 1 \\ \hline
New York & 6 & 0.25 && 5 & 0.1 && 6 & 0.11 & 0 \\ \hline
North Carolina & 1 & 0.04 && 2 & 0.04 && 3 & 0.06 & 1 \\ \hline
Pennsylvania & 1 & 0.04 && 2 & 0.04 && 4 & 0.07 & 1 \\ \hline
Texas & 3 & 0.13 && 11 & 0.23 && 8 & 0.15 & 0 \\ \hline
Virginia & 1 & 0.04 && 3 & 0.06 && 2 & 0.04 & 0 \\ \hline
Washington & 1 & 0.04 && 2 & 0.04 && 2 & 0.04 & 1 \\ \hline
Total & 24 & 0.99 && 48 & 0.99 && 54 & 1.01 & 9 \\ \hline
\end{tabular}
}
\caption{Comparison between realized and proposed center allocations}
\label{tab:comparison_centers}
}
\end{table}

\begin{table}[h]
{\color{black}
\centering
\footnotesize{
\begin{tabular}{l cc c cc c cc c}
\hline
    &   \multicolumn{2}{c}{Realized}    &&  \multicolumn{2}{c}{Proposed}    &&  \multicolumn{2}{c}{Population}   &   \\ \cmidrule{2-3}\cmidrule{5-6}\cmidrule{8-9}
\textbf{State} & \textbf{\#} & \textbf{\%} && \textbf{\#} & \textbf{\%} && \textbf{\#} & \textbf{\%} & \textbf{(R)$\iff$(Pr)} \\ \hline
California & 12 & 0.14 && 53 & 0.1 && 42.5 & 0.22 & 1 \\ \hline
Florida & 12 & 0.14 && 89 & 0.16 && 21.25 & 0.11 & 1 \\ \hline
Illinois & 6 & 0.07 && 13 & 0.02 && 14.167 & 0.07 & 0 \\ \hline
Indiana & 3 & 0.04 && 21 & 0.04 && 7.083 & 0.04 & 1 \\ \hline
Massachusetts & 5 & 0.06 && 12 & 0.02 && 7.083 & 0.04 & 0 \\ \hline
Missouri & 3 & 0.04 && 19 & 0.03 && 3.542 & 0.02 & 1 \\ \hline
New Jersey & 5 & 0.06 && 30 & 0.05 && 7.083 & 0.04 & 1 \\ \hline
New York & 12 & 0.14 && 71 & 0.13 && 21.25 & 0.11 & 1 \\ \hline
North Carolina & 3 & 0.04 && 13 & 0.02 && 10.625 & 0.06 & 1 \\ \hline
Pennsylvania & 6 & 0.07 && 13 & 0.02 && 14.167 & 0.07 & 0 \\ \hline
Texas & 12 & 0.14 && 163 & 0.3 && 28.333 & 0.15 & 0 \\ \hline
Virginia & 5 & 0.06 && 23 & 0.04 && 7.083 & 0.04 & 0 \\ \hline
Washington & 1 & 0.01 && 26 & 0.05 && 7.083 & 0.04 & 0 \\ \hline
Total & 85 & 1.01 && 546 & 0.98 && 191.249 & 1.01 & 7 \\ \hline
\end{tabular}
}
\caption{Comparison between realized and proposed vaccine allocations}
\label{tab:comparison_vaccines}
}
\end{table}

\end{document}